\input harvmac.tex


\let\includefigures=\iftrue
\newfam\black
\includefigures
\input epsf
\def\figin{\epsfcheck\figin}\def\figins{\epsfcheck\figins}
\def\epsfcheck{\ifx\epsfbox\UnDeFiNeD
\message{(NO epsf.tex, FIGURES WILL BE IGNORED)}
\gdef\figin##1{\vskip2in}\gdef\figins##1{\hskip.5in}
\else\message{(FIGURES WILL BE INCLUDED)}%
\gdef\figin##1{##1}\gdef\figins##1{##1}\fi}
\def\DefWarn#1{}
\def\figinsert{\goodbreak\midinsert}
\def\ifig#1#2#3{\DefWarn#1\xdef#1{fig.~\the\figno}
\writedef{#1\leftbracket fig.\noexpand~\the\figno}%
\figinsert\figin{\centerline{#3}}\medskip\centerline{\vbox{\baselineskip12pt
\advance\hsize by -1truein\noindent\footnotefont{\bf Fig.~\the\figno:} #2}}
\bigskip\endinsert\global\advance\figno by1}
\else
\def\ifig#1#2#3{\xdef#1{fig.~\the\figno}
\writedef{#1\leftbracket fig.\noexpand~\the\figno}%
\global\advance\figno by1}
\fi

\def\CA {{\cal A}}

\def\CI {{\cal I}}

\def\CK {{\cal K}}

\def\CM {{\cal M}}
\def\CN {{\cal N}}
\def\CO {{\cal O}}

\def\CV {{\cal V}}
\def\CW {{\cal W}}


\def\p{\partial}
\def\pb{\bar{\partial}}

\def\dd{\nabla}
\def\ddm{\dd_-}
\def\ddp{\dd_+}
\def\ddmd{\dd_\md}
\def\ddpd{\dd_\pd}
\def\ddmm{\dd_\mm}
\def\ddpp{\dd_\pp}
\def\ddpm{\dd_\pm}
\def\ddpmd{\dd_{\dot\pm}}

\def\tilde{\widetilde}
\def\hat{\widehat}
\def\bar{\overline}


\def\p{\partial}
\def\pb{\bar{\partial}}



\def\sqr#1#2{{\vbox{\hrule height.#2pt\hbox{\vrule width
.#2pt height#1pt \kern#1pt\vrule width.#2pt}\hrule height.#2pt}}}
\def\Box{\mathchoice\sqr64\sqr64\sqr{4.2}3\sqr33}

\def\half{{1\over 2}}
\def\a{\alpha}
\def\b{\beta}
\def\g{\gamma}
\def\d{\delta}

\def\p{\partial}
\def\pd{{\dot +}}
\def\md{{\dot -}}

\def\t{{\theta}}
\def\L{\Lambda}

\def\pp{{\mathchoice
              %
          {
              \kern 1pt%
              \raise 1pt
              \vbox{\hrule width5pt height0.4pt depth0pt
                    \kern -2pt
                    \hbox{\kern 2.3pt
                          \vrule width0.4pt height6pt depth0pt
                          }
                    \kern -2pt
                    \hrule width5pt height0.4pt depth0pt}%
                    \kern 1pt
           }
            {
              \kern 1pt%
              \raise 1pt
              \vbox{\hrule width4.3pt height0.4pt depth0pt
                    \kern -1.8pt
                    \hbox{\kern 1.95pt
                          \vrule width0.4pt height5.4pt depth0pt
                          }
                    \kern -1.8pt
                    \hrule width4.3pt height0.4pt depth0pt}%
                    \kern 1pt
            }
            {
              \kern 0.5pt%
              \raise 1pt
              \vbox{\hrule width4.0pt height0.3pt depth0pt
                    \kern -1.9pt  
                    \hbox{\kern 1.85pt
                          \vrule width0.3pt height5.7pt depth0pt
                          }
                    \kern -1.9pt
                    \hrule width4.0pt height0.3pt depth0pt}%
                    \kern 0.5pt
            }
            {
              \kern 0.5pt%
              \raise 1pt
              \vbox{\hrule width3.6pt height0.3pt depth0pt
                    \kern -1.5pt
                    \hbox{\kern 1.65pt
                          \vrule width0.3pt height4.5pt depth0pt
                          }
                    \kern -1.5pt
                    \hrule width3.6pt height0.3pt depth0pt}%
                    \kern 0.5pt
            }
        }}

\def\mm{{\mathchoice
   %
                  %
                       {
                             \kern 1pt
               \raise 1pt    \vbox{\hrule width5pt height0.4pt depth0pt
                                  \kern 2pt
                                  \hrule width5pt height0.4pt depth0pt}
                             \kern 1pt}
                       {
                            \kern 1pt
               \raise 1pt \vbox{\hrule width4.3pt height0.4pt depth0pt
                                  \kern 1.8pt
                                  \hrule width4.3pt height0.4pt depth0pt}
                             \kern 1pt}
                       {
                            \kern 0.5pt
               \raise 1pt
                            \vbox{\hrule width4.0pt height0.3pt depth0pt
                                  \kern 1.9pt
                                  \hrule width4.0pt height0.3pt depth0pt}
                            \kern 1pt}
                       {
                           \kern 0.5pt
             \raise 1pt  \vbox{\hrule width3.6pt height0.3pt depth0pt
                                  \kern 1.5pt
                                  \hrule width3.6pt height0.3pt depth0pt}
                           \kern 0.5pt}
                       }}

\def\ad{{\kern0.5pt
                   \alpha \kern-5.05pt
\raise5.8pt\hbox{$\textstyle.$}\kern 0.5pt}}

\def\bd{{\kern0.5pt
                   \beta \kern-5.05pt \raise5.8pt\hbox{$\textstyle.$}\kern 0.5pt}}

\def\qd{{\kern0.5pt
                   q \kern-5.05pt \raise5.8pt\hbox{$\textstyle.$}\kern 0.5pt}}
\def\Dot#1{{\kern0.5pt
     {#1} \kern-5.05pt \raise5.8pt\hbox{$\textstyle.$}\kern 0.5pt}}

\lref\FMS{D. Friedan, E. Martinec and S. Shenker,
``Conformal Invariance, Supersymmetry and String Theory'',
Nucl. Phys. B271 (1986) 93.}

\lref\N{N. Berkovits, ``Super-Poincar\'{e} Invariant
Koba-Nielsen Formulas for the Superstring'', Phys. Lett. B385 (1996) 109.}

\lref\Na{N. Berkovits, A New Description of the Superstring,
Jorge Swieca School 1995, p. 490.}

\lref\Nat{N. Berkovits, `` Covariant Quantization of the
Green-Schwarz Superstring in a Calabi-Yau Background'', Nucl. Phys. B431,
258 (1994).}

\lref\NV{N. Berkovits and C. Vafa,``$N=4$ Topological Strings''
, Nucl. Phys. B433 (1995) 123.}

\lref\NB{N. Berkovits and B. C. Vallilo, ``Consistency of
Super-Poincar\'{e} Covariant Superstring Tree Amplitudes'', hep-th/0004171.}

\lref\GSW{M. B. Green, J. H. Schwarz and E. Witten, Superstring Theory,
 2 vols., Cambridge University Press (1987)}

\lref\HERM{N. Berkovits, ``Off-Shell Supersymmetry versus Hermiticity
in the Superstring'', Phys. Rev. Letters 77 (1996) 2891, hep-th/9604121.}

\lref\UNIQ{N. Berkovits and C. Vafa, ``On the Uniqueness of String Theory'',
Mod. Phys. Lett. A9 (1994) 653, hep-th/9310129.}

\lref\Eff{N. Berkovits and W. Siegel,
``Superspace Effective Actions for 4D Compactifications of Heterotic
and Type II Superstrings", Nucl. Phys. B462 (1996) 213, hep-th/9510106.}

\lref\Deboer{J. de Boer and K. Skenderis,
``Covariant Computation of the Low Energy Effective Action of
the Heterotic Superstring", Nucl. Phys. B481 (1996) 129, hep-th/9608078.}

\lref\Sixcurved{N. Berkovits, ``Quantization of the Type II Superstring in
a Curved Six-Dimensional Background", Nucl. Phys. B565 (2000) 333,
hep-th/9908041.}
\lref\Windex{E.~Witten, ``Constraints on Supersymmetry Breaking",
Nucl.Phys. {\bf B202} (1982) 253.}
\lref\Witten{E.~Witten, ``Non-Perturbative Superpotentials In
String Theory", Nucl.Phys. {\bf B474} (1996) 343.}
\lref\Wflux{E. Witten, ``On Flux Quantization In M-Theory And
The Effective Action", J. Geom. Phys. {\bf 22} (1997) 1.}
\lref\GVW{S.~Gukov, C.~Vafa and E.~Witten, ``CFT's From
Calabi-Yau Four-folds", hep-th/9906070.}
\lref\BB{K.~Becker and M.~Becker, ``M-Theory on Eight-Manifolds",
Nucl.Phys. {\bf B477} (1996) 155.}
\lref\SVW{S. Sethi, C. Vafa and E. Witten,
``Constraints on Low-Dimensional String Compacti\-fi\-cations",
Nucl. Phys. {\bf B480} (1996) 213.}
\lref\DRS{K.~Dasgupta, G.~Rajesh and S.~Sethi,
``M Theory, Orientifolds and G-Flux", hep-th/9908088.}
\lref\HL{M.~Haack and J.~Louis, ``Aspects of Heterotic/M-Theory
Duality in D=3", hep-th/9908067.}
\lref\HaackL{M.~Haack and J.~Louis, ``M-theory compactified
on Calabi-Yau fourfolds with background flux," hep-th/0103068.}
\lref\HLouis{M.~Haack and J.~Louis, ``Duality in Heterotic Vacua
With Four Supercharges", Nucl.Phys. {\bf B575} (2000) 107;
M.~Haack, J.~Louis and M.~ Marquart, ``Type IIA and Heterotic
String Vacua in D=2," hep-th/0011075.}
\lref\HM{J.A.~Harvey and G.~Moore,
``Superpotentials and Membrane Instantons", hep-th/9907026.}
\lref\Lerche{W.~Lerche, ``Fayet-Iliopoulos Potentials from
Four-Folds", JHEP {\bf 9711} (1997) 004.}
\lref\Mayr{P.~Mayr, ``Mirror Symmetry, N=1 Superpotentials and
Tensionless Strings on Calabi-Yau Four-Folds",
Nucl.Phys. {\bf B494} (1997) 489.}
\lref\Klemm{A.~Klemm, B.~Lian, S.-S.~Roan and S.-T.~Yau,
``Calabi-Yau fourfolds for M- and F-Theory compactifications",
Nucl.Phys. {\bf B518} (1998) 515.}

\lref\GMP{B.R.~Greene, D.R.~Morrison and M.R.~Plesser,
``Mirror Manifolds in Higher Dimension",
Commun.Math.Phys. {\bf 173} (1995) 559.}
\lref\specgeom{B. de~Wit, P.~G.~Lauwers and A. Van~Proeyen,
``Lagrangians of N=2 Supergravity -- Matter Systems",
Nucl.Phys. {\bf B255} (1985) 569.}

\lref\CVfusion{S.~Cecotti and C.~Vafa, ``Topological -- Anti-topological
Fusion", Nucl.Phys. {\bf B367} (1991) 359.}
\lref\CVsigma{S.~Cecotti and C.~Vafa, ``Exact Results for Supersymmetric
Sigma Models", Phys.Rev.Lett. {\bf 68} (1992) 903.}

\lref\Dubrovin{B.~Dubrovin, ``Geometry and Integrability of
Topological-Antitopological Fusion", Commun. Math. Phys. {\bf 152} (1993) 539.}

\lref\NVW{N. Berkovits, C. Vafa and E. Witten, ``Conformal Field Theory of
AdS Background with Ramond-Ramond Flux", JHEP {\bf 9903} (1999) 018,
hep-th/9902098.}

\lref\pures{N. Berkovits, ``Super-Poincare Covariant Quantization of the
Superstring'', JHEP {\bf 04} (2000) 018, hep-th/0001035.}

\lref\GGW{S.J.~ Gates, Jr., S.~ Gukov and E.~ Witten
``Two Two-Dimensional Supergravity Theories from Calabi-Yau Four-Folds",
Nucl.Phys. {\bf B584} (2000) 109.}

\lref\Gukov{S.~ Gukov, ``Solitons, Superpotentials and Calibrations,''
Nucl.Phys. {\bf B574} (2000) 169.}

\lref\KKV{S. Katz, A. Klemm and C. Vafa "Geometric Engineering of
Quantum Field Theories", Nucl.Phys. {\bf B497} (1997) 173.}

\lref\DKL{L.J. Dixon, V. Kaplunovsky and J. Louis, ``On Effective Field
Theories Describing (2,2) Vacua of the Heterotic String",
Nucl.Phys. {\bf B329} (1990) 27.}

\lref\Romans{L.~J.~Romans, ``Massive N=2A Supergravity in Ten
Dimensions", Phys.Lett. {\bf 169B} (1986) 374.}
\lref\PS{J.~Polchinski and A.~Strominger,``New Vacua for Type II
String Theory,'' Phys. Lett. {\bf B388} (1996) 736.}
\lref\deight{E.~Bergshoeff, M.~de Roo, M.~B.~Green, G.~Papadopoulos and
P.~K.~Townsend, ``Duality of Type II 7-branes and 8-branes",
Nucl.Phys. {\bf B470} (1996) 113.}

\lref\CGHS{C. Callan Jr., S.B. Giddings, J.A. Harvey and A. Strominger,
``Evanescent Black Holes", Phys. Rev. D45 (1992) 1005, hep-th/9111056.}.
\lref\gris{M. Grisaru and D. Zanon, ``Dilaton $\beta$-Function on a Kahler
Manifold", Phys. Lett. B184 (1987) 209.}

\lref\KO{M.~ Kato and T.~ Okada, ``D-Branes on Group Manifolds,''
Nucl.Phys. {\bf B499} (1997) 583.}
\lref\Gutperle{M.~ Gutperle and Y.~ Satoh, ``D-branes in Gepner models
and supersymmetry,'' Nucl.Phys. {\bf B543} (1999) 73.}
\lref\Green{M.B.~ Green and M.~ Gutperle, ``Light-cone supersymmetry
and D-branes,'' Nucl.Phys. {\bf B476} (1996) 484.}
\lref\DiV{P.~ Di Vecchia and A.~ Liccardo, ``D branes in string theory, II,''
hep-th/9912275.}

\lref\AS{A.Yu.~ Alekseev and V.~ Schomerus, ``D-branes in the WZW model,"
Phys.Rev. {\bf D60} (1999) 061901, hep-th/9812193.}
\lref\Stanciu{S.~ Stanciu, ``D-branes in group manifolds,"
JHEP {\bf 0001} (2000) 025, hep-th/9909163.}
\lref\FFFG{G.~ Felder, J.~ Frohlich, J.~ Fuchs and C.~ Schweigert
``The geometry of WZW branes,"
J.Geom.Phys. {\bf 34} (2000) 162, hep-th/9909030.}

\lref\OOY{H.~Ooguri, Y.~Oz and Z.~Yin, ``D-Branes on Calabi-Yau
Spaces and Their Mirrors", Nucl.Phys. {\bf B477} (1996) 407.}

\lref\Gradechi{A. M.~ El Gradechi and L. M.~ Nieto, ``Supercoherent States,
Super K\"ahler Geometry and Geometric Quantization,"
Commun.Math.Phys. {\bf 175} (1996) 521, hep-th/9403109.}

\lref\NVW{N. Berkovits, C. Vafa and E. Witten, ``Conformal Field Theory of
AdS Background with Ramond-Ramond Flux", JHEP {\bf 9903} (1999) 018.}

\lref\ADSStwo{N.~Berkovits, M.~Bershadsky, T.~Hauer, S.~Zhukov and
B.~Zwiebach, ``Superstring Theory on AdS$_2 \times S^2$ as a Coset
Supermanifold," Nucl.Phys. {\bf B567} (2000) 61, hep-th/9907200.}

\lref\CDKKTP{P.~Claus, M.~Derix, R.~Kallosh, J.~Kumar, P.K.~Townsend and
A.~Van Proeyen, ``Black Holes and Superconformal Mechanics, ''
Phys.Rev.Lett. {\bf 81} (1998) 4553.}

\lref\CCDFFT{L.~Castellani, A.~Ceresole, R.~D'Auria, S.~Ferrara, P.~Fre' and
M.~Trigiante, ``G/H M-branes and AdS$_{p+2}$ Geometries,''
Nucl.Phys. {\bf B527} (1998) 142.}

\lref\Greene{B.R.~ Greene, K.~ Schalm and G.~ Shiu,
``Warped Compactifications in M and F Theory,''
Nucl.Phys. {\bf B584} (2000) 480.}

\lref\Vafa{C.~Vafa, ``Superstrings and Topological Strings at Large N,''
hep-th/0008142.}

\lref\ADS{J.M.~ Maldacena, "The Large N Limit of Superconformal
        Field Theories and Supergravity", Adv. Theor. Math. Phys.
        {\bf 2} (1998) 231; S.S.~ Gubser, I.~R.~ Klebanov and A.~M.~ Polyakov,
        "Gauge Theory Correlators from Non-Critical String Theory",
        Phys. Lett. {\bf B428} (1998) 105; E.~Witten, "Anti De Sitter
        Space And Holography", Adv. Theor. Math. Phys. {\bf 2} (1998) 253.}

\lref\holography{L.Susskind, "The World as a Hologram,"
J.Math.Phys. {\bf 36} (1995) 6377;
C.R. ~Stephens, G. 't ~Hooft and B.F. ~Whiting, "Black Hole Evaporation
without Information Loss," Class.Quant.Grav. {\bf 11} (1994) 621.}

\lref\talk{S.~Gukov, http://www.cgtp.duke.edu/lecturenotes/2000fall/Gukov.pdf}

\lref\GGWall{S.J.~Gates, Jr., M.T.~Grisaru and M.E.~Wehlau,
``A Study of General 2D, N=2 Matter Coupled to Supergravity in
Superspace", Nucl.Phys. {\bf B460} (1996) 579.}

\lref\GWall{M.T.~Grisaru and M.E.~Wehlau,
``Superspace Measures, Invariant Actions, and Component Projection
Formulae for (2,2) Supergravity'',
Nucl.Phys. {\bf B457} (1995) 219.}

\lref\LVW{W.~Lerche, C.~Vafa and N.P.~Warner, ``Chiral Rings in
N=2 Superconformal Theories", Nucl.Phys. {\bf B324} (1989) 427.}
\lref\Dixon{L.~Dixon, Lectures at the 1987 ICTP summer Workshop
in High Energy Physics and Cosmology.}
\lref\mirrorbook{S.-T.~Yau, editor, {\it Essays on Mirror Manifolds},
International Press, 1992; B.~Greene and S.-T.~Yau, editors,
{\it Mirror Symmetry. II}, International Press, 1997.}

\lref\bars{I.~Bars, ``Free Fields and New Cosets of Current Algebras'',
Phys. Lett. {\bf B255} (1991) 353.}

\lref\WittenWZW{E.~Witten, ``On String Theory and Black Holes,"
Phys.Rev. {\bf D44} (1991) 314.}

\lref\BCOV{M.~ Bershadsky, S.~ Cecotti, H.~ Ooguri and C.~ Vafa,
``Kodaira-Spencer Theory of Gravity and Exact Results for
Quantum String Amplitudes," Commun.Math.Phys. {\bf 165} (1994) 311.}

\lref\AGNT{I.~ Antoniadis, E.~ Gava, K.S.~ Narain and T.R. ~Taylor,
``Topological Amplitudes in Heterotic Superstring Theory,"
Nucl.Phys. {\bf B476} (1996) 133.}

\lref\EFR{S.~ Elitzur, A.~ Forge and E.~ Rabinovici, ``Some global aspects
of string compactifications,'' Nucl. Phys. {\bf B359}, 581 (1991).}

\lref\MSW{G.~Mandal, A.~M.~Sengupta and S.~R.~Wadia,
``Classical solutions of two-dimensional string theory,''
Mod. Phys. Lett. {\bf A6}, 1685 (1991).}

\lref\Mignemi{S.~Mignemi, ``Black hole solutions in generalized
two-dimensional dilation gravity theories,'' Phys. Rev. {\bf D50} (1994) 4733.}

\lref\KKL{M.~O.~Katanaev, W.~Kummer and H.~Liebl, ``On the completeness
of the black hole singularity in 2D dilaton  theories,'' Nucl.Phys. {\bf B486}
(1997) 353.}

\lref\DVV{R.~Dijkgraaf, E.~Verlinde and H.~Verlinde,
Nucl.Phys. {\bf B371} (1992) 269.}

\lref\AO{A.~Achucarro and M.~E.~Ortiz, ``Relating black holes in
two-dimensions and three-dimensions,'' Phys.Rev. {\bf D48} (1993) 3600.}

\lref\MTZ{C.~Martinez, C.~Teitelboim and J.~Zanelli, ``Charged rotating black
hole in three spacetime dimensions,'' Phys.Rev. {\bf D61} (2000) 104013.}

\lref\AT{A.~Achucarro and P.~K.~Townsend, ``A Chern-Simons Action
For Three-Dimensional Anti-De Sitter Supergravity Theories,''
Phys.Lett. {\bf B180} (1986) 89.}

\lref\IT{J.~M.~Izquierdo and P.~K.~Townsend, ``Supersymmetric space-times
in (2+1) adS supergravity models,'' Class.Quant.Grav. {\bf 12} (1995) 895.}

\lref\Henneaux{O.~ Coussaert and M.~ Henneaux, ``Supersymmetry of
the 2+1 black holes,'' Phys.Rev.Lett. {\bf 72} (1994) 183.}
\lref\HH{J.B.~Hartle and S.W.~Hawking, ``Solutions of the Einstein-Maxwell
Equations with Many Black Holes,'' Commun.Math.Phys. {\bf 26} (1972) 87.}
\lref\MMS{J.~Maldacena, J.~Michelson and A.~Strominger,
``Anti-de Sitter Fragmentation,'' JHEP {\bf 9902} (1999) 011.}

\lref\Mann{R.B.~Mann, ``Conservation Laws and 2D Black Holes in Dilaton
Gravity,'' Phys.Rev. {\bf D47} (1993) 4438.}

\lref\LVA{H. ~Liebl, D.V.~ Vassilevich and
S.~ Alexandrov, ``Hawking radiation and
masses in generalized
dilaton theories,'' Class.Quant.Grav. {\bf 14} (1997) 889.}

\lref\Polyakov{A.M.~Polyakov, Phys.Lett. {bf 103B} (1981) 207.}

\lref\Lowe{D.A.~ Lowe, ``Semiclassical Approach to Black Hole Evaporation,''
Phys.Rev. {\bf D47} (1993) 2446.}

\lref\LL{D.A.~ Lowe and M.~ O'Loughlin, ``Nonsingular Black Hole
Evaporation and ``Stable'' Remnants,'' Phys.Rev. {\bf D48} (1993) 3735.}

\lref\Hawking{S. W. Hawking, ``Evaporation of Two Dimensional Black Holes,''
Phys.Rev.Lett. {\bf 69} (1992) 406.}

\lref\Banks{T.~Banks and M.~O'Loughlin, ``Nonsingular Lagrangians for
Two Dimensional Black Holes,'' Phys.Rev. {\bf D48} (1993) 698.}

\lref\ST{A. Strominger and S.P. Trivedi, ``Information Consumption
by Reissner-Nordstrom Black Holes,'' Phys.Rev. {\bf D48} (1993) 5778.}

\lref\ENTANGLE{L.~Susskind and J.~Uglum, ``Black Hole Entropy in Canonical
Quantum Gravity and Superstring Theory,'' Phys.Rev. {\bf D50} (1994) 2700;
T.M.~Fiola, J.~Preskill, A.~Strominger and S.P.~Trivedi,
``Black hole thermodynamics and information loss in two dimensions,''
Phys.Rev. {\bf D50} (1994) 3987;
T.~Jacobson, ``Black Hole Entropy and Induced Gravity,'' gr-qc/9404039.}

\lref\HMS{S.~Hawking, J.~Maldacena and A.~Strominger, ``DeSitter entropy,
quantum entanglement and ADS/CFT,'' JHEP {\bf 0105} (2001) 001.}

\lref\BanksN{T.~Banks, ``Cosmological Breaking of Supersymmetry?,''
hep-th/0007146.}

\lref\Wittenvard{E. Witten, ``String Theory Dynamics In Various
Dimensions," Nucl.Phys. {\bf B443} (1995) 85.}

\lref\BKORP{E.~Bergshoeff, R.~Kallosh, T.~Ortin, D.~Roest and
A.~Van Proeyen, ``New Formulations of D=10 Supersymmetry
and D8-O8 Domain Walls,'' hep-th/0103233.}

\lref\toappear{N.~Berkovits and S.~Gukov, work in progress.}

\lref\Ortin{B.~Janssen, P.~Meessen and T.~Ortin, ``The D8-Brane
Tied up: String and Brane Solutions in Massive Type IIA Supergravity,''
Phys.Lett. {\bf B453} (1999) 229.}

\lref\ParkS{Y.~Park, A.~Strominger, ``Supersymmetry and Positive
Energy in Classical and Quantum Two-Dimensional Dilaton Gravity,''
Phys.Rev. {\bf D47} (1993) 1569.}

\lref\KKK{V.~Kazakov, I.~Kostov and D.~Kutasov, ``A Matrix Model
for the Two Dimensional Black Hole,'' hep-th/0101011.}

\lref\Katanaev{M.O.~Katanaev, ``Global Solutions in Gravity.
Lorentzian signature,'' Proc.Steklov Inst.Math. {\bf 228} (2000) 158.}

\lref\Katanaeveff{M.O.~ Katanaev, ``Effective action for scalar
fields in two-dimensional gravity,'' gr-qc/0101033.}

\lref\Filippov{A.T.~Filippov,
``Exact Solutions of $(1+1)$-Dimensional Dilaton
Gravity Coupled to Matter,'' Mod.Phys.Lett. {\bf A11} (1996) 1691;
``Integrable 1+1 dimensional gravity models,''
Int.J.Mod.Phys. {\bf A12} (1997) 13.}

\lref\Zaslavskii{O.B.~Zaslavskii, ``Exactly solvable models of
two-dimensional dilaton gravity and quantum eternal black holes,''
Phys. Rev. {\bf D59} (1999) 084013; 
``Semi-infinite throats at finite temperature and static solutions in exactly
solvable models of 2d dilaton gravity,'' Phys.Lett. {\bf B459} (1999) 105.}

\lref\Odintsov{S.~Nojiri, S.D.~Odintsov, `` Quantum dilatonic gravity
in d = 2,4 and 5 dimensions,'' Int.J.Mod.Phys. {\bf A16} (2001) 1015.}

\lref\Solodukhin{S.N.~Solodukhin, ``Two-dimensional Quantum-Corrected
Eternal Black Hole,'' Phys. Rev. {\bf D53} (1996) 824.}

\lref\Sorkintoadd{E.~Sorkin, T.~Piran, ``Formation and Evaporation
of Charged Black Holes,'' Phys.Rev. {\bf D63} (2001) 124024.}

\lref\Tseytlintoadd{J.~Russo, A.A.~Tseytlin, ``Scalar-Tensor Quantum
Gravity in Two Dimensions,'' Nucl.Phys. {\bf B382} (1992) 259.}

\lref\Nappitoadd{M.D.~McGuigan, C.R.~Nappi, S.A.~Yost,
``Charged Black Holes in Two-Dimensional String Theory,''
Nucl. Phys. {\bf B375} (1992) 421.}


\Title{ \vbox{\baselineskip12pt
\hbox{hep-th/0107140}
\hbox{IFT-P.052/2001}
\hbox{ITEP-TH-72/00}
\hbox{CITUSC/00-067}
\hbox{CALT-68-2312}}}
{{\vbox{\centerline{Superstrings in 2D Backgrounds with R-R Flux}
\smallskip
\centerline{and New Extremal Black Holes}
}}}
\centerline{Nathan Berkovits,$^{\spadesuit}$\foot{E-Mail: nberkovi@ift.unesp.br}
Sergei Gukov,$^{\clubsuit}$\foot{E-Mail: gukov@feynman.princeton.edu}
and Brenno Carlini Vallilo$^{\spadesuit}$\foot{E-Mail: vallilo@ift.unesp.br}}
\bigskip
\centerline{\it $^{\spadesuit}$
Instituto de F\'\i sica Te\'orica, Universidade Estadual
Paulista}
\centerline{\it Rua Pamplona 145, 01405-900, S\~ao Paulo, SP, Brasil}
\smallskip
\centerline{\it $^{\clubsuit}$Joseph Henry Laboratories,
Princeton University, Princeton, NJ 08544,}
\centerline{\it California Institute of Technology, Pasadena, CA 91125, USA}

\vskip .2in

The hybrid formalism is used to quantize the superstring compactified
to two-dimensional target-space in a manifestly spacetime supersymmetric
manner. A quantizable sigma model action is then constructed for
the Type II superstring in curved two-dimensional supergravity backgrounds
which can include Ramond-Ramond flux. Such curved backgrounds include
Calabi-Yau four-fold compactifications with Ramond-Ramond flux,
and new extremal black hole solutions in two-dimensional dilaton
supergravity theory. These black hole solutions are a natural
generalization of the CGHS model and might be possible to describe
using a supergroup version of the SL(2,R)/U(1) WZW model.
We also study some dynamical aspects of the new black holes,
such as formation and evaporation.


\Date{July 2001}


\newsec{Introduction}

Since Ramond-Ramond vertex operators in the Ramond-Neveu-Schwarz (RNS)
formalism
are represented by spin fields which mix the matter
and ghost sector, it is
difficult to use this formalism to quantize the superstring
in Ramond-Ramond (R-R) backgrounds.
Although the covariant Green-Schwarz (GS)
formalism can classically describe R-R backgrounds, this
formalism has quantization
problems even in a flat background.

Over the last seven years, a hybrid formalism has been developed
which combines advantages of the RNS and GS formalisms. Like the RNS
formalism, the hybrid formalism has a free-field action in a flat
background so it is easy to quantize. And like the GS formalism,
the worldsheet variables include the target superspace coordinates
$(x,\t)$ so target-space supersymmetry is manifest and R-R backgrounds
are easy to describe. Furthermore, the hybrid formalism is related
by a field redefinition to the $N=1\to N=2$
embedding of the RNS formalism \UNIQ\NV, and the resulting critical
$\hat c=2$ $N=2$ superconformal field theory (scft) splits into a
$\hat c= (d-6)/2$ $N=2$ scft for the target space and a
$\hat c= (10-d)/2$ $N=2$ scft for the internal compactification
manifold. The $N=2$ scft for the internal
compactification manifold is related to the analogous RNS $\hat c=(10-d)/2$
$N=2$ scft by twisting the U(1)-charged compactification-dependent
worldsheet fields.
This hybrid formalism has been previously
developed for compactification to $d=6$
on Calabi-Yau (CY) two-folds \NV\NVW\Sixcurved\
and for compactification to $d=4$ on
CY three-folds \Nat\Eff.
It will be developed here for compactification to $d=2$
on CY four-folds.

Even though in two dimensions IIA R-R fields have no propagating
degrees of freedom, their zero-modes play a very important role ---
they induce a back reaction on the graviton generating a negative
cosmological constant in the effective two-dimensional theory.
Therefore, in addition to describing CY four-fold
compactifications with R-R flux, the hybrid formalism developed
here can be used
to study $D=2$ dilaton supergravity with an additional R-R term,
which is a natural generalization of the CGHS model.
This $D=2$ supergravity theory has new extremal black hole
solutions which will be shown to have interesting properties.

The paper is organized as follows:
In section 2, we develop the hybrid formalism for
superstring compactification on CY four-folds
to flat two-dimensional Minkowski
space. For heterotic compactifications, $N=(2,0)$
spacetime supersymmetry is manifest, while for IIA (or IIB)
compactifications, $N=(2,2)$ (or $N=(4,0)$) spacetime supersymmetry
is manifest. After defining the free-field action and worldsheet
$N=2$ superconformal generators in the hybrid formalism, vertex operators
for physical massless states are explicitly constructed.
For the Type II superstring, these massless states include four
R-R field strengths which are independent of the compactification.
These universal combinations of R-R fields
will be identified with four possible types of F-terms in
the effective $\CN=2$ theory: superpotential, twisted superpotential,
and their complex conjugates.
As in \Vafa, the hybrid formalism can be used to compute ``topological''
contributions to these superpotentials.

In section 3, the
superstring action and worldsheet $N=2$ superconformal generators
are generalized for compactifications to
a curved $d=2$ target-superspace background. The
resulting sigma model action is
manifestly target-space super-reparameterization invariant and can be used
to quantize the superstring in two-dimensional backgrounds with
R-R flux. The sigma model action contains a Fradkin-Tseytlin term
which couples a spacetime superfield containing
the dilaton to the worldsheet $N=(2,2)$
supercurvature. As in the $d=4$ and $d=6$ hybrid formalisms \Eff\Sixcurved,
this coupling is possible since superspace chirality in spacetime is
related to superspace chirality on the worldsheet.
The background superfields appearing in the sigma model are then
used to construct a low-energy superpace effective action for the
$N=(2,2)$ supergravity theory.

In section 4, we study the bosonic contribution to the
low-energy effective action of
the two-dimensional supergravity theory.
A particular feature of this action is a
(cosmological) constant term, independent of the dilaton,
which is induced by the background R-R flux.
The action without this term was extensively studied
in the literature in the early 90s.
In particular, it was shown that it has classical
black hole solutions \refs{\EFR,\MSW,\WittenWZW},
which evaporate \CGHS, and eventually become singular \Hawking,
in quantum theory coupled to matter.
We show that with an extra term corresponding
to the R-R flux, the theory admits a large family of novel
black hole solutions including extremal black holes,
analogous to those in \Nappitoadd.
We conjecture that the extremal solutions represent stable
$1/2$ BPS states of the two-dimensional supergravity theory
and support this conjecture by calculating the mass
of non-singular black hole solutions
and demonstrating that it satisfies a BPS-like inequality.
Then we examine some dynamical aspects of the new black holes,
in the classical and semiclassical approximations.
Namely, we study
formation and evaporation of the near-extremal black holes and argue
that 
the new extremal
black holes do not emit Hawking radiation.
%

Finally, in section 5 we conclude with a discussion of
open problems which include finding a supergroup sigma model
to describe the superstring quantized on the black hole
solutions of section 4. Such a sigma model could be useful
for studying quantum corrections to the solutions, in a manner
analogous to the SL(2,R)/U(1) gauged WZW sigma model description \WittenWZW\
of the CGHS black holes \refs{\MSW, \CGHS}.


\newsec{Hybrid Formalism for Compactification to Two Flat Dimensions}

In this section we describe the basic methods of the hybrid
formalism and how to apply them in the case of superstring
compactification to a flat two-dimensional background.
The first step will be to find a field redefinition from RNS variables
to hybrid variables which allows $d=2$ spacetime supersymmetry to
be made manifest. The next step will be to construct worldsheet
$N=2$ superconformal generators in the hybrid formalism using the $N=1\to N=2$
embedding of the RNS superstring. The final step will be to use these
$N=2$ superconformal generators to define
physical vertex operators in the hybrid formalism.

\subsec{Field redefinition from RNS variables}

The basic variables
of the RNS description of the superstring  are $(x^m, \psi^m)$
and the (bosonized)
superconformal ghosts $(b,c,\beta=e^{-\phi}\p\xi,\gamma=e^{\phi}\eta)$
which gauge-fix the $N=1$ supergravity of the worldsheet.
For superstring compactifications which preserve at least $N=2$ supersymmetry
in flat two-dimensional space-time, these variables factorize into a
set describing the flat space $(x^\pp,x^\mm,\psi^\pp,\psi^\mm)$
plus the superconformal ghosts and a $\hat c=4$ $N=2$ scft
associated with the eight-dimensional compactification manifold,
{\it e.g.} a Calabi-Yau four-fold.
The action has the form (for simplicity
we will discuss only the left moving sector in this section):
\eqn\FLAT{ S_{flat}= \int d^2z [ \p x^\mm \bar\p x^\pp
- \psi^\pp \pb\psi^\mm] +
S_{CY} + S_{ghosts},}
where $S_{ghosts}$ is the action for the superconformal ghosts
and $S_{CY}$ is the sigma model action for the compactification manifold,
which we assume to be a Calabi-Yau four-fold for the sake of concreteness.
The explicit form of these actions will not be needed.

As discussed in \FMS, spacetime supersymmetry can be studied defining the zero
mode of the vertex
operator of the massless fermionic state at zero momentum to be the
supersymmetry charge.
Since there are only two non-compact dimensions, the spin fields
take a very simple form,
$\Sigma^\alpha =e^{\pm\half\sigma}$ where $\sigma$ is the
chiral boson which
bosonize the $(\psi^\pp,\psi^\mm)$ variables. The GSO projection tells us
that we must choose one
chirality for $\Sigma^\alpha$ and we will choose $\Sigma^-$.
In the $-\half$ picture,
there are two supercharges, $q_+ =\oint e^{-\half \phi}\Sigma^-
e^{+\half H_{CY}}$ and
$q_\pd =\oint e^{-\half \phi}\Sigma^- e^{-\half H_{CY}}$, where
$J_{CY}=\p H_{CY}$ is the worldsheet current contained in the $N=2$
superconformal algebra of the CY sigma model. The fact that the supersymmetry
operator carries
picture tell us that supersymmetry in the RNS model is well defined
only for on-shell states. This can also be seen verifying that these operators
satisfy the $d=2$ $N=2$ supersymmetry algebra up to a picture change.

Nevertheless, in two spacetime dimensions it is
possible to define supersymmetry
off-shell using the
methods of \Nat\NV\NVW . To do this, we define $q_\pd$ in the $-\half$
picture as above
but define its complex conjugated $q_+$ in the $\half$ picture as
\eqn\Q{ q_+ = \oint( b\eta e^{{3\over 2} \phi}\Sigma^- e^{\half H_{CY}} -
e^{\half\phi}: [ \p x^\mm\Sigma^+  + (G^+_{CY} + G^-_{CY})\Sigma^- 
]e^{\half H_{CY}}:) ,}
where $G^+_{CY}$
and $G^-_{CY}$ are superconformal generators of the CY sigma model.
For this to be consistent,
we must change the definition of complex conjugation. This can
be done following \HERM, and will not be described explicitly here.
The two generators $q_\pd$ and $q_+$ satisfy the desired supersymmetry algebra.

Now we define two superspace
variables $\t^+$ and $\t^\pd$ that when commuted with
$q_+$ and $q_\pd$ respectively give the identity operator:
\eqn\THETAS{  \t^+ = c\xi
e^{-{3\over 2}\phi}\Sigma^+ e^{-\half H_{CY}} , \quad
\t^\pd = e^{\half\phi}\Sigma^+ e^{\half H_{CY}}.}
Note that
these operators have
conformal weight zero and transform as spacetime spinors. In
addition, we define conformal weight one operators conjugate to $\t$'s:
\eqn\MOMENTA{ p_+ = b\eta e^{{3\over 2}\phi}\Sigma^- e^{\half H_{CY}}, \quad
p_\pd = e^{-\half\phi}\Sigma^- e^{-\half H_{CY}} .}

We will treat these fields
as the fundamental ones. Remembering the initial number of
degrees of freedom, $(x^\pp, x^\mm, \psi^\pp, \psi^\mm, b,c,\beta,\gamma)$,
we see that there are two independent
chiral bosons remaining to be defined. These two chiral bosons,
$\omega$ and $\rho$, will be defined as
\eqn\RHOOMEGA{ \omega = \half(\phi -\kappa + \sigma -\chi), \quad
\rho  = 2\phi + \half ( H_{CY} -\chi - 3 \kappa), }
where $\p\chi=cb$
and $\p\kappa=\xi\eta$. These chiral bosons have the OPE's:
\eqn\BOPES{ \omega(z)\omega(w)
\to \half \log (z-w), \quad \rho(z)\rho(w)\to -\half \log (z-w).}

The set of hybrid
variables is now complete. The next step is to make a chiral $U(1)$
rotation on the compactification
fields in order they do not have singular OPE's
with the hybrid variables due to the $H_{CY}$
dependence. If a compactification field
$\Phi_{CY}$ has $U(1)$ charge $n$, we will define a new field
$\Phi_{GS}=e^{n(\phi -\kappa)}\Phi_{CY} $. This transformation changes the
compactification dependent
superconformal generators by a similarity transformation
$\CO_{GS}= e^{\oint(\phi
- \kappa)J_{CY}}\CO_{CY}e^{-\oint(\phi - \kappa)J_{CY}}$. In
particular, $G^\pm_{GS} = e^{\pm(\phi-\kappa)} G^\pm_{CY}$
and $J_{GS}=J_{CY} + 4(\p\phi -\p\kappa )$.

\subsec{Worldsheet $N=2$ superconformal generators}

As discussed in \UNIQ ,
the generators  $(T=T_{matter}^{RNS} + T_{ghost}^{RNS}, G^+=j_{BRST},
G^-=b, J=cb +\xi\eta)$ of the
combined matter ghost
system form an $N=2$ twisted superconformal
algebra with central
charge $\hat c=2$. These generators can be untwisted by defining
$T \to T-\half \p J$ and rewritten using the hybrid
variables.
Using the definition of $p_+$, $\omega$ and $\rho$,
it is easy to see
that $J= -2\p\rho + J_{GS}$ and
$G^- =b= p_+e^{\omega - \rho}.$

$G^+=j_{BRST}$ can be determined in a similar way (e.g.
$-\gamma^2 b = p_\pd e^{\omega + \rho}$) and can be checked by
requiring consistency of the
algebra. $T$ is uniquely
determined by conformal weights of the fundamental fields and
requiring $G^+$ and $G^-$
to have conformal weight $3\over 2$. The complete
set of untwisted generators is:

\eqn\GENE{T = \Pi\pp \Pi\mm - d_+\p\t^+ - d_\pd\p\t^\pd + \p\omega\p\omega
-\p\rho\p\rho -\p^2\omega + T_{GS},}
$$G^- = d_+e^{\omega - \rho},$$
$$G^+ = (\Pi_\mm d_\pd + 2 \p \omega \p\t^+ -\half \p^2\t^+) e^{-\omega +\rho}
+ d_{\pd} e^{\omega+\rho} + G_{GS}^+ + \t^+ T_{GS}e^{-\omega +\rho} +
G^-_{GS}\t^+p_\pd e^{2\rho},$$
$$J= -2\p\rho + J_{GS},$$
where
\eqn\SUSYOPE{\Pi_\mm = \p x_\mm + \t^+ \p\t^\pd  ,\quad \Pi_\pp = \p x_\pp,}
$$d_+ = p_+ , \quad d_\pd = p_\pd +\t^+ \p x\pp,$$
are supersymmetric operators.

This can be
expressed in a more elegant form by doing a similarity transformation
in the hybrid part
$$\CO_{hybrid}\to e^{\oint (\t^+ e^{\rho -\omega}G^-_{GS}-
\half \p x_\pp \t^+\t^\pd)}\CO_{hybrid}
e^{-\oint (\t^+ e^{\rho -\omega}G^-_{GS}
-\half \p x_\pp \t^+\t^\pd)}.$$
Note that it does not change the conformal
weight or the charge of the operators, puts a term $G^-_{GS}$ in $G^-$
and removes the terms $\t^+ T_{GS}
e^{-\omega +\rho} + G^-_{GS}\t^+p_\pd e^{2\rho}$ in $G^+$.
This transformation also allows us to write the $d=2$
supersymmetric operators and supercharges as:
\eqn\SUSYOPEfinal{\Pi_\mm = \p x_\mm +\half( \t^+ \p\t^\pd +\t^\pd \p\t^+)
,\quad \Pi_\pp = \p x_\pp,}
$$d_+ = p_+ + \half \t^\pd\p x_\pp, \quad d_\pd = p_\pd +\half\t^+ \p x\pp,$$
\eqn\SUSYCHARGfinal{q_+ =\int( p_+ -\half\t^\pd\p x_\pp), \quad q_\pd = \int
(p_\pd -
\half\t^+\p x_\pp)}
and the
superconformal generators are
\eqn\gen{T = \Pi_\pp \Pi_\mm - d_+\p\t^+ - d_\pd \p\t^\pd + \p \omega\p \omega
 - \p^2 \omega - \p^2 \rho + T_{GS},}
$$G^- = d_+ e^{w-\rho}
 + G_{GS}^-,$$
$$G^+ =  e^{\oint \Pi_\mm e^{-2w}}(d_{\pd}e^{\omega+\rho})
e^{-\oint \Pi_\mm e^{-2w}}  + G_{GS}^+,$$
$$J = -2 \p\rho + J_{GS},$$

The worldsheet action for these hybrid variables is
\eqn\hyb{ S_{hybrid}= \int d^2z [ \p x^\mm \bar\p x^\pp
- p_+ \pb\t^+ - p_\pd \pb\t^\pd] +
S_C + S_B,}
where $S_C$ is the action for the new $\hat c=4$ scft and
$S_B$ is the action for the chiral bosons $\rho$ and $w$.
Like its RNS counterpart of \FLAT, \hyb\ is a free-field action if
one ignores the scft for the compactification. But $S_{hybrid}$
has the advantage over \FLAT\ that the hybrid variables transform
covariantly under the spacetime supersymmetry transformations generated
by \SUSYCHARGfinal.
In the closed string case there are also hybrid variables
corresponding to the right movers.
The difference between type IIA and IIB is the choice of the
chiralities of the spinor variables. Note that the right mover
chiral boson $\bar \omega$ has different $SO(1,1)$ charges in
the two cases.

\subsec{Definition of physical states}

With this superconformal algebra, one can introduce $N=2$
superconformal ghosts and carry
out a standard BRST procedure to define physical states \Nat .
But since we have already included the RNS
matter and ghost fields to define the hybrid variables, the N=2
superconformal ghosts play a trivial role and will be ignored
in the following discussion.

For open superstrings,
physical vertex operators can be defined as real, chiral,
or antichiral $N=2$ primary fields constructed from the hybrid
variables.\foot{An alternative way to
define physical states in the hybrid formalism
is to use the $N=4$
topological method \NV\ in which the $\hat c=2$ $N=2$ generators
are extended to a set of $\hat c=2$ $N=4$ generators. Although
this method is useful for constructing gauge-invariant actions or
computing scattering amplitudes, it is unnecessary if one only
wants to construct physical states.} Real primary fields $\Phi$ carry
zero $U(1)$ charge and zero conformal weight and satisfy
$G_n^+ \Phi= G_n^-\Phi=0$ for $n\geq 0$ where
$\CO_n\CA$ means the coefficient of the pole of order $n+h_\CO$ in
the OPE of
$\CO$ and $\CA$.
Chiral primary fields $\Phi^c$ carry
$+1$ $U(1)$ charge and $+\half$ conformal weight and satisfy
$G_{n-1}^+ \Phi^c= G_n^-\Phi^c=0$ for $n\geq 0$, and
antichiral primary fields $\Phi^a$ carry
$-1$ $U(1)$ charge and $+\half$ conformal weight and satisfy
$G_n^+ \Phi^a= G_{n-1}^-\Phi^a=0$ for $n\geq 0$.
These primary fields are defined up to gauge transformations
$\d \Phi = G^+_{-n}\L_n + G^-_{-n}\bar\L_n$ for $n>0$ which preserve
the above conditions.

For real primary fields, the integrated open superstring vertex
operator is $V= \int dz G^-_{-\half} G^+_{-\half} \Phi$, while for
chiral and antichiral primary fields,
the integrated open superstring vertex
operator is $V= \int dz G^-_{-\half} \Phi^c$ and
$V= \int dz G^+_{-\half} \Phi^a$. Physical closed superstring vertex operators
can be defined in the usual manner as left-right products of open
superstring vertex operators. For example, for real primary fields,
the integrated closed superstring vertex operator is
\eqn\IV{ \CV = \int d^2z \bar G^-_{-\half}
\bar G^+_{-\half} G^-_{-\half} G^+_{-\half} \Phi  .}

\subsec{Compactification-independent massless states}

We will first discuss massless open string
states independent of the compactification.
The most general massless compactification-independent operator is

$$\Phi(x,\t)= V + U_\mm e^{-2w}, $$
where $\Phi$
in the massless case depends only on the zero modes of $x$ and $\t$, i.e.
there are no derivatives on them. Using the real primary field conditions
that $T_0\Phi = G^\pm_\half\Phi=0$
and the OPE's
$$x^\pp(y) x^\mm(z) \to 2\log|y-z|,\quad
w(y) w(z) \to \half\log(y-z),\quad
\rho(y) \rho(z) \to -\half\log(y-z),$$
$$\t^\pd(y) p_\pd (z) \to (y-z)^{-1},\quad
\t^+(y) p_+ (z) \to (y-z)^{-1},$$
we get the following superfield equations:
\eqn\SMEE{\Box V = \Box U_\mm =0}
\eqn\SME{\ddp U_\mm =0,}
\eqn\SMe{\ddpd( U_\mm - \dd_\mm V) =0 ,}
where
$$\ddmm = \p_\mm, \quad \ddpp = \p_\pp,$$
$$\ddp = \p_+ + \half \t^\pd\p_\pp, \quad \ddpd = \p_\pd + \half\t^+\p_\pp.$$

The residual gauge invariance from
$\d\Phi = G^+_{-\half}\L + G^-_{-\half}\bar\L$
are generated by $\L =  e^{-w - \rho}
\L_\md ,\quad \bar\L = e^{-w +3\rho - H_{CY}}\L_- ,$
with the gauge transformations:
$$\d V = \ddp \L_- + \ddpd \L_\md ,$$
$$\d U_\mm = \ddmm\ddpd\L_\md  ,$$
satisfying $\ddmm\ddp\ddpd \L_\md = \ddmm\ddpd\ddp \L_- =0$.

The solution of the equations  is
\eqn\ram{\Phi= \t^+\phi^-(x^\mm) +\t^\pd\phi^{\md} (x^\mm)
+ \t^\pd\p_\mm\phi^\md (x^\mm) e^{-2\omega}.}
The fields that are not eliminated by the equations of motion are
pure gauge. Only the Ramond sector contains physical degrees
of freedom which are
two left-moving fermions, agreeing with the standard light-cone
analysis.

In the case of closed Type II
superstrings, the situation is unusual. Since the states of
the closed
string are just the product of the open string states, the physical degrees
of freedom are in the R-R sector.
For Type IIB we will get four chiral bosons (depending on
$x^\mm$), but for Type IIA we will have only constant
modes since the product
of two chiral fermions of opposite chiralities ( the left moving contribution
will depend on $x^\mm$
and the right moving will depend on $x^\pp$)  must be constant.
Since constant R-R fluxes play an important role in CY four-fold
compactifications, it will be interesting to discuss these
compactification-independent states in more detail.

\subsec{Type IIA R-R vertex operators}

When $\p_\mm\phi^\md$ and
$\p_\mm\phi^-$ are constants in \ram, the integrated vertex operators
$V=\int dz G^-_{-\half} G^+_{-\half}\Phi$ associated with these
open superstring states are the spacetime supersymmetry generators
$q_+$ and $q_\pd$ of \SUSYCHARGfinal. The Type IIA vertex operators
for the four constant R-R field-strengths are therefore
\eqn\rrv{ V_{cc} = \int d^2z j_+j_-, \quad
 V_{aa} = \int d^2z j_\pd j_\md, \quad
V_{ca} = \int d^2z j_+ j_\md, \quad
V_{ac}=
 \int d^2z j_\pd j_-, }
where $q_\pm=\oint j_\pm$ and $q_{\dot\pm}=\oint j_{\dot\pm}$
are the four Type IIA supercharges.

In terms of the original 32
ten-dimensional IIA spacetime-supersymmetry generators
$q_\a = q_{\pm\pm\pm\pm\pm}$,
\eqn\orig{q_+ = q_{+++++},\quad
q_\pd = q_{+----},\quad q_-= q_{-++++},\quad
q_\md = q_{-----}.}
The vertex operator described by $\int d^2 z j_\a j_\b$
couples to the string coupling constant $e^\varphi$ times the
ten-dimensional IIA R-R field strengths
in the combination
\eqn\combr{F_{\a\b}=
\d_{\a\b} F^{(0)} + \half\g^{MN}_{\a\b} F_{MN}^{(2)} +  {1\over{24}}
\g^{MNPQ}_{\a\b} F_{MNPQ}^{(4)}}
where $M=0$ to 9.
Since they carry $+4$ and $-4$ CY $U(1)$ charge,
the compactification-independent R-R vertex operators
$V_{cc}$ and
$V_{aa}$ are easily seen to
couple to the holomorphic and anti-holomorphic four-form R-R
field strengths $\Omega^{jklm} F^{(4)}_{jklm}$ and
$\Omega^{\bar j\bar k\bar l\bar m}F^{(4)}_{\bar j\bar k\bar l\bar m}$
where $j,k,l,m=1$ to 4 denote the complex Calabi-Yau directions,
and $\Omega$ is the covariantly constant $(4,0)$-form.
In other words, $V_{cc}$ and $V_{aa}$ couple to the following
components of the Ramond-Ramond field strengths:
\eqn\vertexidb{F_{+-} = W, \quad F_{\pd \md} = \bar W}
where, following the notations of \Gukov, we introduced:
\eqn\wone{ W = {1 \over 2 \pi} \int_X \Omega \wedge F^{(4)}}

Similarly, one can check that
the two other compactification-independent R-R vertex operators,
$V_{ca}$ and
$V_{ac}$,
couple to $e^\varphi$ times two specific linear combinations
of the field strengths
$F^{(0)}$, $F^{(2)}_{01}$, $F^{(2)}_{j\bar j}$, $F^{(4)}_{01 j\bar j}$
and $F^{(4)}_{jk\bar j\bar k}$.
The precise linear combination of field strengths which couple to
$V_{ca}$ and $V_{ac}$ can
be derived using the mirror map of the CY manifold which
maps the holomorphic $F^{(4)}_{jklm}$ and the anti-holomorphic
$F^{(4)}_{\bar j\bar k\bar l\bar m}$
into these linear combinations.
The idea is that the conformal field theory associated
with the mirror variety $\tilde X$ is physically equivalent
to the CFT associated with the original CY four-fold $X$.
Essentially, the two are related by a change of
notations \refs{\LVW,\Dixon},
which exchanges $- \leftrightarrow \md$.
This transformation implies a non-trivial isomorphism of
the cohomology rings of $X$ and its mirror $\tilde X$, such that:
\eqn\hmirror{ h^{p,q} (X) = h^{4-p,q} (\tilde X)}
Therefore, we claim that R-R flux corresponding, say, to $V_{ca}$
is given by the image of $W$ on $\tilde X$ under
the mirror map, which we now describe in more detail.

Mirror symmetry is usually stated as a map from the moduli
space $\CM_c (X)$ of complex structure of $X$ to the moduli
space of K\"ahler structure $\CM_K (\tilde X)$ of the mirror
manifold $\tilde X$ \mirrorbook.
Following \GMP, we denote by $D$ a connection on the holomorphic
bundle over the complex structure moduli space $\CM_c (X)$.
Starting with the holomorphic 4-form $\Omega$, we can
act on it with $D$ to generate the cohomology elements
of $\oplus_{k=0}^4 H^{k,4-k} (X)$. The space spanned by
these elements is called the ``horizontal primary subspace'':
$$H_H (X) \subset \oplus_{k=0}^4 H^{k,4-k} (X).$$

Similarly, one can consider wedge products
of the elements $\CO_i \in H^{1,1} (\tilde X)$
in the cohomology of the mirror variety $\tilde X$.
The subspace of $\oplus_{k=0}^4 H^{k,k} (\tilde X)$
spanned by these elements is called the ``vertical primary subspace'':
$$H_V (\tilde X) \subset \oplus_{k=0}^4 H^{k,k} (\tilde X).$$

Mirror symmetry for Calabi-Yau four-folds states that
these two spaces we have just introduced,
$H_H (X)$ and $H_V (\tilde X)$, are isomorphic.
We denote this isomorphism as:
\eqn\mirror{\pi_m \colon H_H (\tilde X) \longrightarrow H_V (X) }

Since the RR vertex operators $V_{ca}$ and $V_{ac}$ are related
to $V_{cc}$ and $V_{aa}$ by mirror map in the world-sheet conformal
field theory, the corresponding Ramond-Ramond fields must be
also related by the mirror map \mirror. Therefore, the RR vertex
operators $V_{ca}$ and $V_{ac}$ represent the following
compactification-independent components of the Ramond-Ramond fields:
\eqn\vertexid{\eqalign{F_{+ \md} &= \tilde W_X = \pi_m (W_{\tilde X}), \cr
F_{\pd -} &= \tilde {\bar W}_X = \pi_m (\bar W_{\tilde X}) }}
Using the mirror map \mirror\ we can explicitly write down
the expression for $\tilde W$ \OOY:
\eqn\wtwo{\tilde W = {1 \over 2 \pi} \int_X e^{\CK} \wedge
\Big( F^{(0)} + i F^{(2)} + F^{(4)} + i F^{(6)} + F^{(8)} \Big) }
where $\CK$ is the K\"ahler form on the CY space $X$, $F^{(6)} = * F^{(4)}$
and $F^{(8)} = * F^{(2)}$.
%

\subsec{Compactification-dependent massless states}

For open string compactification dependent states, the massless states
are given by  chiral and antichiral primary vertex operators
described by $\Phi^c = ( V_i + e^{-2w}U_{\mm i })\Psi_c^i$ and
$\Phi^a = ( \bar V_{\bar i} +
e^{-2w}\bar U_{\mm \bar i })\Psi_a^{\bar i}$ operators
where $\Psi^i_c$ and
$\bar\Psi^{\bar i}_{a}$ are chiral and antichiral left-moving
primary operators
of the CY scft. The chiral and antichiral primary conditions,
$G^-_{\half}\Phi^c = G^+_{-\half}\Phi^c = 0$ and
$G^+_{\half}\Phi^a = G^-_{-\half}\Phi^a = 0$, imply
the superfield equations
$$U_{\mm i } = -\ddmm V_i, \quad \ddp V_i = 0 ,
\quad \ddmm \ddpd V_i = 0, $$
$$\bar U_{\mm \bar i } = 0, \quad \ddmm\ddp
\bar V_{\bar i} = 0, \quad \ddpd \bar V_{\bar i} = 0.$$

The solution of these equations is:
$$ V_i = a(x^\mm,x^\pp) + \t^\pd b(x^\pp)
-\half \t^+\t^\pd \p_\pp a(x^\mm,x^\pp) $$
$$ \bar V_{\bar i}= \bar a(x^\mm,x^\pp) + \t^+\bar b(x^\pp)
+\half \t^+\t^\pd \p_\pp \bar a(x^\mm,x^\pp)  $$
which is an N=(2,0) $d=2$ scalar multiplet containing two real bosons
and two left-moving fermions.

The closed string compactification-dependent states
are obtained by taking the left-right product of open string
states. For each $(c,c)$ or $(c,a)$ CY modulus, the Type IIB superstring
has a massless N=(4,0) $d=2$ multiplet containing two real bosons,
four left-moving fermions, and two left-moving R-R bosons.
For each $(c,c)$ or $(c,a)$ CY modulus, the Type IIA superstring
has a massless N=(2,2) $d=2$ scalar multiplet containing two real bosons,
two left-moving fermions, two right-moving fermions, and two constant
R-R bosons. As will be discussed in the following subsection, the R-R
bosons in these multiplets play an interesting role in the low-energy
superpotential terms.

\subsec{ Computation of superpotential terms}

In the large volume limit, one can perform Kaluza-Klein reduction
of type IIA supergravity on a Calabi-Yau space $X$ with RR flux in order
to find the effective action for $2D$ bosonic fields.
The bosonic part of the effective Lagrangian of massive Type IIA
supergravity in string frame looks as follows \refs{\Wittenvard,\BKORP}:
\eqn\iiaaction{L_{IIA} = \int d^{10} x \sqrt{- g} \Big[ e^{-2 \varphi}
\big( R + 4 (\nabla \varphi)^2 - {1 \over 12} H^2 \big) -
\sum_p {1 \over 2 \cdot (2p)!} \big( F^{(2p)} \big)^2 + \ldots \Big] }
Here the dots stand for higher derivative $R^4$ and other terms.
Even though we don't write these terms explicitly,
they play an important role in the anomaly cancellation condition.
Taking into account the back reaction effects of the RR flux, such as
warping of the space-time geometry, from the dimensional reduction
of \iiaaction\ one can obtain the bosonic part of the effective $\CN=(2,2)$
two-dimensional dilaton supergravity interacting with matter \GGW.
In particular, one finds the effective superpotential \wone\ induced by
the RR flux (along with the twisted chiral superpotential \wtwo)
\refs{\Gukov,\GVW,\HaackL}:
\eqn\wwone{ W = {1 \over 2 \pi} \int_X \Omega \wedge F^{(4)}}

This expression for the effective superpotential is protected
by supersymmetry and can be also derived from (topological)
string amplitudes. The
manifestly supersymmetric hybrid approach turns out to be
particularly useful for this sort of calculations \NV.
Here we illustrate it for the case when $F^{(4)}$-flux is
a linear combination of $(3,1)$ and $(1,3)$ forms, {\it cf.} \Vafa.
The space of $(3,1)$ forms, $H^{3,1} (X)$, is generated by
$\omega_i^{(3,1)} = D_{\phi_i} \Omega$ where $\{ \phi_i \}$ are
{\it special} coordinates on the complex structure moduli space $\CM_c (X)$.
Therefore, we can take R-R flux in the following
form\foot{This notation means that $F^{(m,n)}$ contains $m$
holomorphic and $n$ anti-holomorphic indices in the CY directions.}:
\eqn\ganzats{F^{(3,1)} = \sum_i n_i ( D_{\phi_i} \Omega) }

Recall, that for a generic CY four-fold compactification
of type IIA superstring, complex structure moduli give rise to
chiral superspace multiplets in the two-dimensional effective
field theory ({\it i.e.} massless $\CN=(2,2)$ superfields
$\Phi_i (x^{\mu}, \theta^+,  \theta^- )$ satisfying
$\nabla_{\dot +} \Phi_i = \nabla_{\dot -} \Phi =0$).
{}From the Kaluza-Klein reduction one finds the superspace
action of these modes \GGW:
\eqn\kact{S^{(2)} = \int d^2 x d^4 \theta e^{- K(\Phi_i, \bar \Phi_i )} =
\int d^2 x d^4 \theta \Big( \int_X \Omega \wedge \bar \Omega \Big)}
where we ignored interaction with gravity.
{}From the supersymmetry structure of the vertex operators
for the fields $\Phi_i$ it follows that the auxiliary field in $\Phi_i$
corresponds to a Ramond-Ramond flux, so that we have:
\eqn\phistr{\Phi_i = \phi_i + (\theta)^2 n_i + \ldots}
Turning on a R-R flux of the form \ganzats\ means giving
a vev to the auxiliary fields in $\Phi_i$. Therefore, apart from
the usual kinetic action for $\phi_i$, integrating over two
$\theta$ variables in \kact\ we also get an F-term:
$$
\int d^2 x d^2 \theta \Big( \sum_i n_i \int_X
D_{\phi_i} \Omega \wedge \bar \Omega \Big) + c.c.
$$
As in \Vafa, this expression is exactly the same as the F-term that one
finds after substituting \ganzats\ into superpotential \wwone.
In a similar way, one should be able to reproduce other relations
between D-terms and F-terms in the two-dimensonal effective
action from (topological) string amplitudes controlled by
supersymmetry\foot{An example of such relations is a generalization
of the special geometry constraints, which by analogy with
\refs{\DKL,\specgeom} can be formulated as a set of equations for
the metric $G_{i \bar j}$ on the complex structure moduli space
of $X$ and holomorphic four-point correlation function $\kappa_{ijkl}$ \GMP:
$$
R_{i\bar j k \bar l}- G_{i \bar j} G_{k \bar l}-
G_{i \bar l} G_{k \bar j} = e^K B^{pq \bar s \bar t}
\kappa_{pqik}  {\bar \kappa_{\bar s \bar t \bar j \bar l}}
$$
Here $R_{i\bar j k \bar l}= - \bar\partial_{\bar j}\big(G_{k\bar
m}\partial_i G^{\bar m n}\big) G_{n \bar l}$ is the curvature,
and $(B^{-1})_{pq \bar s \bar t} =
e^{-K} \big( G_{p \bar t} G_{q \bar s}
+ G_{q \bar t} G_{p \bar s} - R_{p \bar t q \bar s} \big)$,
provided that $h^{2,2}_{prim} = h^{3,1} (h^{3,1} + 1)/2$.
In type IIA compactifications on Calabi-Yau four-folds the use
of such relations was first pointed out in \Lerche.}.

\newsec{Hybrid Formalism in Curved $d=2$ Target-Space Background}

The usual method for constructing the
superstring action in a curved target-space background is to add the
on-shell
massless vertex operator to the action in a flat background, and
covariantize with respect to target-space (super-)reparameterizations.
This procedure is not very useful for two-dimensional target-space
backgrounds where almost all massless vertex operators are trivial
on-shell. Nevertheless, it is easy to guess the correct superstring action in
a curved two-dimensional target-superspace by simply changing
the range of the superspace indices in the hybrid actions
for the superstring in a curved four-dimensional \Eff\ or six-dimensional
\Sixcurved\ background.

As in the four and six-dimensional cases, the sigma model action in
a two-dimensional background splits into a ``classical'' term and
a Fradkin-Tseytlin term which couples the spacetime dilaton to
the worldsheet $N=(2,2)$ supercurvature. After describing these two
terms, it will be argued that worldsheet $N=(2,2)$ superconformal
invariance of this action implies the appropriate low-energy superspace
equations of motion for the background superfields.

\subsec{Classical sigma model action}

For the Type IIA superstring compactified on a Calabi-Yau four-fold,
the ``classical'' part of the worldsheet action is\foot{
For convenience, we will only discuss here
the sigma model action for the Type IIA superstring. However, it is
easy to generalize this action to the heterotic or Type IIB superstrings
as was discussed in \Eff.}
\eqn\two{S ={1\over {\a'}}\int dz d\bar z
(\Pi_\pp \bar\Pi_\mm + B_{AB} \Pi^A \bar\Pi^B}
$$ + d_+ \bar\Pi^+
 +\bar d_\pd \bar\Pi^\pd + d_- \Pi^-
+ \bar d_\md \Pi^\md $$
$$+ d_+ d_- Y_{cc}
+ \bar d_\pd \bar d_\md Y_{aa}
+ d_+ \bar  d_\md  Y_{ca}
+ \bar d_\pd  d_-  Y_{ac} ) $$
$$+ S_C + S_B$$
where
$\Pi^A = E_M^A \p Z^M$,
$\bar\Pi^A = E_M^A \bar\p Z^M$,
$A,B$ range over the tangent
superspace indices $(\pp,\mm,+,\pd,-,\md)$,
$M$ ranges over the curved superspace indices, $Z^M=(x,\t,\bar\t)$ are
the $d=2$ $N=(2,2)$ target superspace coordinates,
and $S_C$ and $S_B$ are actions for the
compactification manifold and for the chiral bosons.
Worldsheet fields carrying
$-$ or $\md$ spinor indices come from the right-moving sector
of the IIA superstring and those carrying
$+$ or $\pd$ spinor indices come from the left-moving sector.
Since the worldsheet variables $p_\pm$ and $\bar p_{\dot\pm}$
never appear explicitly in \two, $d_\pm$ and $\bar d_{\dot\pm}$ can
be treated as independent worldsheet variables.
Note that
the background superfields $[Y_{cc},Y_{aa}, Y_{ca}, Y_{ac}]$ need
to be included in \two\ to reproduce the R-R vertex operators of \rrv.

Although $S_C$ is the same action as in a flat $d=2$ background, $S_B$
needs to be modified since
the chiral and antichiral
bosons $(\omega,\rho,\bar \omega,\bar\rho)$ are no longer
free worldsheet fields.
Since they carry Lorentz weight and $R$-charge, they couple to the
$SO(1,1)$ Lorentz and $U(1)\times U(1)$ connections
of the target-space supergravity theory which will be called
$A^L_M$,
$A_M^1$ and $A_M^2$. From their Lorentz and $R$-charges, one
finds that the action $S_B$ in a curved background is
\eqn\curvedboson{S_B = S_B^{flat} + \int dz d\bar z [
\bar\p Z^M (\p \omega A^L_M + \p\rho A_M^1) +
\p Z^M (-\bar\p \bar \omega A^L_M + \bar\p\bar\rho A_M^2) ]}
where $S_B^{flat}$ is the action for free chiral and antichiral bosons.

The background superfields appearing in \two\ and \curvedboson\ are the
$d=2$ $N=(2,2)$ target-space super-vierbein $E^A_M(Z)$, the
Lorentz and $U(1)\times U(1)$ connections $[A^L_M(Z),A^1_M(Z),A^2_M(Z)]$,
the tensor two-form superfield $B_{AB}(Z)$, and the four scalar superfields
[$Y_{cc}(Z)$, $Y_{aa}(Z)$, $Y_{ca}(Z)$, $Y_{ac}(Z)$] whose lowest components
are the string coupling constant times
the four compactification-independent R-R field strengths discussed
in subsection 2.5.
As will be discussed in subsection 3.4, worldsheet $N=(2,2)$
superconformal invariance of the sigma model is expected to relate
these background superfields to each other and impose their torsion
constraints and low-energy
equations of motion.

Note that \two\ is manifestly target-space
super-reparameterization invariant and reduces to the free-field action
of \hyb\ when the background superfields take their flat values.
Although \two\ has been written in worldsheet conformal gauge, it
is easy to write it in a manifestly worldsheet reparameterization invariant
manner by coupling appropriately to the worldsheet vierbein. However,
since the worldsheet variables do not transform linearly under
worldsheet $N=(2,2)$ supersymmetry (i.e. under commutation with
$G^\pm_{-\half}$ and $\bar G^\pm_{-\half}$), it is not known how
to write \two\ in a manifestly worldsheet super-reparameterization
invariant manner.

\subsec{Fradkin-Tseytlin term}

Because the expectation value of the spacetime dilaton is
the logarithm of the
string coupling constant, one expects a Fradkin-Tseytlin term
which couples this spacetime field to the worldsheet curvature.
As in the four and six-dimensional hybrid formalisms, this term
is constructed by coupling chiral and twisted-chiral spacetime
superfields containing the dilaton
to chiral and twisted-chiral worldsheet superfields which
describe the worldsheet $N=(2,2)$ supercurvature.
This will be done in a consistent manner
since worldsheet superspace chirality is related to target-superspace
chirality in the hybrid formalism.
So the
Fradkin-Tseytlin term couples the target-space and worldsheet
$N=2$ $d=2$ supergravities to each other.

In the $U(1)\times U(1)$ version of $N=2$ supergravity, the
worldsheet supercurvature is described by a chiral
and twisted-chiral superfield, $\Sigma_{cc}$ and
$\Sigma_{ca}$, and their complex conjugates $\Sigma_{aa}$ and
$\Sigma_{ac}$, satisfying
\eqn\sigc{D_\pd\Sigma_{cc}=
D_\md\Sigma_{cc}=0,
\quad D_\pd\Sigma_{ca}=
D_-\Sigma_{ca}=0,}
$$
D_+\Sigma_{aa}=
D_-\Sigma_{aa}=0,
\quad D_+\Sigma_{ac}=
D_\md\Sigma_{ac}=0.$$
$D_\pm$ and $D_{\dot\pm}$ are the worldsheet $N=(2,2)$
supersymmetric derivatives which satisfy
$\{D_+,D_\pd\}=\p_z$ and
$\{D_-,D_\md\}=\bar\p_{\bar z}$ on a flat worldsheet.
The ordinary
worldsheet curvature is given by the lowest component
of $Re(D_+ D_-\Sigma_{cc} + D_+ D_{\dot -}\Sigma_{ca})$, and the
two worldsheet $U(1)$ field strengths are given by the lowest
component of $Im(
D_+ D_-\Sigma_{cc} + D_+ D_{\dot -}\Sigma_{ca})$ and
$Im(
D_+ D_-\Sigma_{cc} + D_\pd D_-\Sigma_{ac})$.

The Fradkin-Tseytlin term is constructed by coupling these
worldsheet superfields to the chiral and twisted-chiral spacetime
superfields, $\Phi_{cc}$ and $\Phi_{ca}$, and their complex conjugates
$\Phi_{aa}$ and $\Phi_{ac}$. These spacetime superfields satisfy
\eqn\phic{\ddpd\Phi_{cc}=
\ddmd\Phi_{cc}=0,
\quad \ddpd\Phi_{ca}=
\ddm\Phi_{ca}=0,}
$$
\ddp\Phi_{aa}=
\ddm\Phi_{aa}=0,
\quad \ddp\Phi_{ac}=
\ddmd\Phi_{ac}=0$$
where $\ddpm$ and $\ddpmd$ are the spacetime $N=(2,2)$
supersymmetric derivatives satisfying
$\{\ddp,\ddpd\}=\ddpp$ and
$\{\ddm,\ddmd\}=\ddmm$.
The lowest component of $Re(\Phi_{cc}+\Phi_{ca})$ is the
spacetime dilaton, while the lowest component of
$Im(
\Phi_{cc} + \Phi_{ca})$ and
$Im(
\Phi_{cc} + \Phi_{ac})$ are spacetime $U(1)\times U(1)$ compensators
which will be discussed in subsection 3.4.

As in the four and six-dimensional versions of the hybrid formalism,
the Fradkin-Tseytlin term is
defined as
\eqn\ft{S_{FT}= \int d^2 z [G^-_{-\half} \bar G^-_{-\half} (\hat\Phi_{cc}
 \Sigma_{cc})
+ G^-_{-\half} \bar G^+_{-\half} (\hat\Phi_{ca} \Sigma_{ca})}
$$
+G^+_{-\half} \bar G^+_{-\half} (\hat\Phi_{aa} \Sigma_{aa})
+ G^+_{-\half} \bar G^-_{-\half} (\hat\Phi_{ac} \Sigma_{ac})]$$
where $G^\pm_{-\half}\hat\Phi$ or $\bar G^\pm_{-\half} \hat\Phi$ is
defined using the OPE's of the worldsheet fields and
$G^\pm_{-\half}\Sigma$ or $\bar G^\pm_{-\half}\Sigma$ is defined as
the appropriate worldsheet supersymmetric derivative acting on $\Sigma$.
In other words, treating the fermionic
$N=2$ superconformal generators as worldsheet supersymmetry derivatives,
one defines
\eqn\wsp{G^+_{-\half}\Sigma = D_+\Sigma,\quad
G^-_{-\half}\Sigma = D_\pd\Sigma,\quad
\bar G^+_{-\half}\Sigma = D_- \Sigma,\quad
\bar G^-_{-\half}\Sigma = D_\md \Sigma,}
since
$+/-$ and $\pd/\md$ target-space and worldsheet
spinor indices correspond to unbarred/barred (or left/right moving)
superconformal generators.
The hatted $\Phi$ superfield is related to the unhatted
$\Phi$ superfield by
\eqn\defphihat{
\hat\Phi_{cc} = (1 +  e^{-2\omega}\ddpp +
e^{-2\bar \omega}\ddmm + e^{-2\omega-2\bar \omega}\ddpp\ddmm)
\Phi_{cc}, \quad \hat\Phi_{aa} = \Phi_{aa}}
$$\hat\Phi_{ca} = (1 +  e^{-2\omega}\ddpp )\Phi_{ca},\quad
\hat\Phi_{ac} = (1 +  e^{-2\bar\omega}\ddmm )\Phi_{ac}.$$
The $e^{-\omega}$ factors are included in \defphihat\ so that
\phic\ implies that
$\hat\Phi_{cc}$ satisfies $G^+_{-\half}\hat\Phi_c =
\bar G^+_{-\half}\hat\Phi_c=0$ and is therefore a chiral worldsheet field.
Similarly, \phic\ implies that
$\hat\Phi_{ca}$ is a twisted-chiral worldsheet field,
$\hat\Phi_{aa}$ is an antichiral worldsheet field,
and
$\hat\Phi_{ac}$ is a twisted-antichiral worldsheet field.
So the Fradkin-Tseytlin term of \ft\ is worldsheet supersymmetric as desired.
Furthermore,
one can check that it contains the usual term $\int dz d\bar z \varphi r$
where $\varphi$ is the spacetime dilaton and $r$ is the worldsheet curvature.

\subsec{Worldsheet $N=(2,2)$ superconformal invariance}

Although the action of \two\ is classically
worldsheet conformally invariant,
$\a'$ quantum corrections from sigma model loops will spoil
this symmetry. After adding contributions from the Fradkin-Tseytlin
term of \ft\ and putting the background on-shell, quantum conformal
invariance is expected to be restored. However, unlike the bosonic
string sigma model, vanishing of the $\b$-functions is not strong
enough to imply the complete set of low-energy equations of motion
for the background. This can be seen from a linearized expansion
of the sigma model action into the free-field action in a flat
background plus the massless closed superstring vertex operator.
One can easily see that conformal invariance only implies that
the vertex operator is annihilated by
$T_0$ and $\bar T_0$, but does not imply the primary field conditions
associated with $G^\pm$ and $\bar G^\pm$.

One therefore expects that quantum worldsheet $N=(2,2)$ superconformal
invariance of the sigma model must be imposed in order to obtain the
complete set of torsion constraints and low-energy equations of motion
for the background superfields. Since worldsheet supersymmetry is
not manifest in the hybrid formalism, the simplest way to verify
worldsheet $N=(2,2)$ superconformal invariance is to check that
the worldsheet $N=(2,2)$ superconformal generators in a curved
background satisfy the correct OPE's.

For the hybrid formalism in a four-dimensional background, these
OPE's were explicitly computed at tree-level (for heterotic and Type II
compactifications) and one-loop (for heterotic compactifications),
and shown to imply the correct torsion constraints and low-energy
equations of motion for the background superfields. Since the
sigma model action in a four-dimensional background closely resembles
the action of \two\ and \ft, one expects this will
also be true in a two-dimensional background.

In a curved background described by \two\ and \ft,
the worldsheet $N=(2,2)$ superconformal generators
are defined as
\eqn\genc{T = \Pi_\pp \Pi_\mm - d_+\Pi^+ - d_\pd \Pi^\pd +
(\p \omega - A^L_M\p Z^M)^2
-(\p \rho - A_M^1 \p Z^M)^2 }
$$
- \p (\p \omega - A^L_M\p Z^M)
 + T_{GS}
 +  \a'\p^2 (\hat\Phi_{cc} +\hat\Phi_{aa} +
\hat\Phi_{ca} +\hat\Phi_{ac} ) $$
$$G^- = d_+ e^{w-\rho} + G_{GS}^-
 + G^-_{-\half} \p(\hat\Phi_{cc} +\hat\Phi_{ca}), $$
$$G^+ = e^{\oint \Pi_\mm e^{-2\omega}}(d_\pd e^{w+\rho})
e^{-\oint \Pi_\mm e^{-2\omega}} + G_{GS}^+
 + G^+_{-\half} \p(\hat\Phi_{aa} +\hat\Phi_{ac}), $$
$$J = -2 \p\rho + J_{GS}
+\p(
\hat\Phi_{cc} -\hat\Phi_{aa} +
\hat\Phi_{ca} -\hat\Phi_{ac}),$$
\vskip 10pt
\eqn\genbc{\bar T = \bar\Pi_\pp \bar\Pi_\mm - d_-\bar\Pi^- -
d_\md \bar\Pi^\md +
(\bar\p \bar\omega + A^L_M\bar\p Z^M)^2
-(\bar\p \rho - A_M^2 \p Z^M)^2 }
$$
- \bar\p (\bar\p \omega + A^L_M\bar\p Z^M)
 + \bar T_{GS}
 +  \a'\bar\p^2 (\hat\Phi_{cc} +\hat\Phi_{aa} +
\hat\Phi_{ca} +\hat\Phi_{ac} ) $$
$$\bar G^- = d_- e^{\bar w-\bar\rho} + \bar G_{GS}^-
 +\bar G^-_{-\half} \bar\p(\hat\Phi_{cc} +\hat\Phi_{ac}), $$
$$\bar G^+ = e^{\oint \bar\Pi_\pp e^{-2\bar\omega}}(d_\md
e^{\bar w+\bar \rho})
e^{-\oint \bar\Pi_\pp e^{-2\bar\omega}} +\bar G_{GS}^+
 +\bar G^+_{-\half} \bar\p(\hat\Phi_{aa} +\hat\Phi_{ca}), $$
$$\bar J = -2 \bar\p\bar\rho +\bar J_{GS}
+\bar\p(
\hat\Phi_{cc} -\hat\Phi_{aa} -
\hat\Phi_{ca} +\hat\Phi_{ac}).$$
Note that the Fradkin-Tseytlin contribution to the
constraints of \genc\ are analogous to those discussed in
\Eff\ and \Deboer\ for
the four-dimensional background.

\subsec{Torsion constraints and conformal compensators}

Although no explicit calculations will be done here, it is
easy to generalize the results of \Eff\ to determine
the superspace torsion constraints which come from imposing tree-level
worldsheet superconformal invariance of the sigma model.
Using the results of \Eff\ and reducing the target-space
indices from
four-dimensional superspace to two-dimensional superspace,
one obtains the superspace torsion constraints
\eqn\torsion{ Y_{cc} = T_{\pd \mm}^- = T_{\md\pp}^+,\quad
Y_{aa} = T_{+ \mm}^\md = T_{-\pp}^\pd,}
$$ Y_{ca} = T_{\pd \mm}^\md = T_{-\pp}^+,\quad
Y_{ac} = T_{+ \mm}^- = T_{\md\pp}^\pd,$$
$$T_{+\pd}^\pp = T_{-\md}^\mm= H_{\pp -\md} = H_{\mm +\pd}=1,$$
where $T_{AB}^C$ is the torsion defined using
$[\dd_A,\dd_B] = T_{AB}^C \dd_C + ...$ and
$H_{ABC} = \dd_{[A} B_{BC]} + T_{[AB}^D B_{C]D}$ is the tensor
field strength.

The first two lines of \torsion\ imply that the R-R field strengths
are contained in the supergravity multiplet and
the third line
of \torsion\ implies that spacetime conformal invariance of the
background superfields has been broken by fixing $H_{\pp -\md}$
to be a constant. Note that Bianchi identities imply that
all components of $H_{ABC}$ are determined by a scalar superfield
$G$ satisfying  $H_{\pp -\md}=
H_{\mm +\pd} = G$. However, if the target-space
theory is made conformally invariant by including conformal
compensator superfields,
one can fix conformal invariance by gauging $G=1$.

As in the four and six-dimensional
hybrid formalisms, exponentials
of the chiral and twisted-chiral spacetime superfields
appearing in the Fradkin-Tseytlin term of \ft\ will play the role of the
$U(1)\times U(1)$ conformal compensator superfields. These
compensator superfields are convenient
for constructing superspace actions for supergravity \GWall.
Under conformal and $U(1)\times U(1)$ $R$ transformations parameterized
by the superfields $L$, $R^1$ and $R^2$, the logarithm of the
compensator superfields will be defined to transform as
\eqn\conft{\d\Phi_{cc} =i(R_1 + R_2) + L , \quad
\d\Phi_{ca} = i(R_1 - R_2) + L ,}
$$\d\Phi_{ac}  = i(-R_1 + R_2)+ L, \quad
\d\Phi_{aa} = i(-R_1 - R_2)+ L.$$

Since the torsion constraints of \torsion\ break
conformal invariance by fixing
$H_{\pp -\md}$ to a constant, the real part of $\Phi_{cc}+\Phi_{ca}$
is an independent degree of freedom whose lowest component is the spacetime
dilaton. But since spacetime $U(1)\times U(1)$ $R$ symmetry is unbroken
by the torsion constraints, the imaginary part of $\Phi_{cc}+\Phi_{ca}$
and $\Phi_{cc}+\Phi_{ac}$ are gauge degrees of freedom which can be
fixed to zero.

\subsec{Superspace effective action}

As in the four-dimensional hybrid formalism, the low-energy
superspace effective action can be constructed in a spacetime
conformally invariant manner from the superfields
$E_A^M$, $G$ and $e^\Phi$ which were discussed in
the previous subsection. In Einstein gauge, one breaks conformal invariance
by gauge-fixing $e^\Phi=1$
which produces a component action of the form
$S= \int d^D x [ \sqrt{g} R + ...].$ But in string gauge, one breaks
conformal invariance by gauge-fixing $G=1$ which produces a component
action of the form
$S= \int d^D x e^{-2\varphi}[ \sqrt{g} R + ...].$ The low-energy
effective action coming from the sigma model is in string gauge since
the dilaton coupling in the
Fradkin-Tseytlin term implies that all NS-NS terms appear with a factor
of $e^{-2\varphi}$.

As in the four-dimensional
and ten-dimensional Type II effective actions, the kinetic
term for the R-R gauge fields, $(d A_{R-R})^2$,
appears with no $e^{-2\varphi}$ factor.
The explanation of this fact \Eff\ is that the R-R field strength $d A_{R-R}$
appears in the sigma model through the $(\theta)^2$ component
of the compensator superfields $e^{-\Phi}$, i.e.
\eqn\compex{e^{-\Phi_{cc}} = e^{-\phi_{cc}} + \t^+\t^- F_{+-} + ...,\quad
e^{-\Phi_{aa}} = e^{-\phi_{aa}} + \t^\pd\t^\md F_{\dot +\dot -} + ...,}
$$
e^{-\Phi_{ca}} = e^{-\phi_{ca}} + \t^+\t^\md F_{+\dot -} + ...,\quad
e^{-\Phi_{ac}} = e^{-\phi_{ac}} + \t^\pd\t^- F_{\dot + -} + ...,$$
where $F_{\alpha \beta}$ is defined in
\combr. Of the four scalars appearing in \compex,
the dilaton $\varphi$ is identified with $\half Re(\phi_{cc} +\phi_{ca}).$
$Im(\phi_{cc})$ and $Im(\phi_{ca})$ can be gauged away using the
$U(1)\times U(1)$ gauge invariance, and the remaining scalar,
$Re(\phi_{cc} -\phi_{ca})$, will vanish on-shell.
Note that in four dimensions,
\compex\ implies that $e^{-\Phi}$ must
satisfy certain Bianchi identities, but these identities are absent
in two dimensions since the field strengths are scalars.
In two dimensions, the field strength superfield for a vector
multiplet is obtained by simply dualizing the auxiliary scalar(s)
in a chiral or twisted-chiral superfield \GGWall.

Ignoring the CY moduli, the
low-energy superspace action is expected to be
\eqn\superspaceaction{S=
\int d^2 x\int d\t^+ d\t^- d\t^\pd d\t^\md ~(sdet E) ~
(e^{-\Phi_{cc}-\Phi_{aa}} -
 e^{-\Phi_{ca}-\Phi_{ac}}).}
As in four dimensions,
$\int d^4\theta e^{-2\Phi}$ produces $F^2$ terms with no $e^{-2\varphi}$
factor as can be seen from the component expansion of \compex. 
And the R-R component
fields which appear
in the lowest component of
$[Y_{cc}, Y_{ca},
Y_{ac}, Y_{aa}]$ of \torsion\
are actually auxiliary fields $y_{\alpha\beta}$
which are related to the R-R field strengths $F_{\alpha\beta}$
by the equations of motion $y_{\a\b}=e^\varphi F_{\a\b}$.
Up to proportionality constants, the  compactification-independent
bosonic contribution
to the low-energy effective action is
\eqn\lowenergy{S= \int d^2 x \sqrt{-g} [ e^{-2\varphi} (R +4
\ddpp\varphi\ddmm\varphi
+y_{+-}y_{\pd\md}+y_{\pd -}y_{+ \md}) }
$$ -  e^{-\varphi}(y_{+-}F_{\pd\md}+y_{\pd\md}F_{+-}
+y_{+\md}F_{\pd -} + y_{\pd -}F_{+\md})$$
$$ - F_{+-}F_{\pd \md} - F_{+\md}F_{\pd -} + d ~
(e^{-\phi_{cc}-\phi_{aa}}
-  e^{-\phi_{ca}-\phi_{ac}}) ],$$
where $\varphi=\half Re (\phi_{cc}+\phi_{ca})$ and $d(x)$
is an auxiliary field contained in the $\t^4$ component of $sdet E$.
After integrating out $y_{\alpha\beta}$ and $d$,
\eqn\efftwo{S= \int d^2 x \sqrt{-g}[ e^{-2\varphi}
(R +4 \ddpp\varphi\ddmm\varphi )
 -2 F_{+-}F_{\pd \md} - 2 F_{+\md}F_{\pd -} ].}

In the next section, we shall consider adding to the action
a cosmological term proportional to $e^{-2\varphi}$, i.e.
$c\int d^2 x \sqrt{-g} e^{-2\varphi}$, which arises when the total
matter and ghost central charges do not cancel.
Such a term can be supersymmetrized by adding the chiral and twisted-chiral
F-terms
\eqn\ctc{S' =\int d^2 x [c_{\pd\md}\int d\t^+d\t^- e^{-2\Phi_{cc}} +
 c_{+-} \int d\t^\pd d\t^\md e^{-2\Phi_{aa}}}
$$ +
 c_{\pd -} \int d\t^+d\t^\md e^{-2\Phi_{ca}} +
 c_{+ \md} \int d\t^\pd d\t^- e^{-2\Phi_{ac}}] $$
to the action of \superspaceaction. Using the component analysis of
\compex, the bosonic contribution to $S'$ is
\eqn\ctcomp{S' = 2\int d^2 x \sqrt{g} e^{-\varphi}
(c_{\pd\md} F_{+-} +
c_{+-} F_{\pd\md} +
c_{\pd -} F_{+\md} +
c_{+\md} F_{\pd -} ).}

Note that
in the presence of $S'$, the equations of motion for the R-R field
strengths are $\p_m ( F_{\a\b} - c_{\a\b} e^{-\varphi})=0.$
Plugging these auxiliary equations of motion into the action
$S+S'$,
one obtains
\eqn\twodaction{
S_{eff} = S + S' = \int d^{2} x \sqrt{- g} \Big[ e^{-2 \varphi}
\big( R + 4 (\nabla \varphi)^2 + c  \big) + \Lambda \Big] }
where $\Lambda= - 2 \big( \vert W \vert^2 + \vert \tilde W \vert^2 \big)$
and 
$c = 2(\vert c_{+ -}\vert^2 + \vert c _{+ \md}\vert^2)$ are constants, and
\eqn\wdefi{W = F_{+-} - c_{+-} e^{-\varphi}, \quad
\bar W = F_{\pd\md} - c_{\pd\md} e^{-\varphi}, \quad
\tilde W = F_{+\md} - c_{+\md} e^{-\varphi}, \quad
\tilde{\bar W} = F_{\pd -} - c_{\pd -} e^{-\varphi}.}
Note that spacetime supersymmetry implies that $c\geq 0$ and
$\Lambda\leq 0$.


\newsec{New Two-Dimensional Black Holes}

In this section we study
the effective action of superstrings in two-dimensional backgrounds
with Ramond-Ramond flux and ``excess'' central charge \twodaction:
\eqn\tdaction{S_{2d} = \int d^{2} x \sqrt{- g} \Big[ e^{-2 \varphi}
\big( R + 4 (\nabla \varphi)^2 + c \big) + \Lambda \Big]}
Here both the ``excess central charge'' $c$ and
the ``cosmological constant'' $\Lambda$ have dimension
mass squared since we absorbed a factor of $\alpha'$
in the definition of $c$.

As we explained above,
the generalized dilaton gravity action \tdaction\
is very natural from
the string theory point of view, since it includes both the central
charge term $c$ and the cosmological constant term $\Lambda$
corresponding to background Ramond-Ramond fields.
In the special limiting cases, corresponding to $\Lambda=0$ and to $c=0$,
our action \tdaction\ reduces, correspondingly, to the CGHS model \CGHS\
also inspired by string theory,
and to the dilaton gravity model studied by Mignemi \Mignemi.
General dilaton gravity Lagrangians of the form \tdaction\ have been considered
in the literature, see {\it e.g.}
\refs{\Nappitoadd, \Banks, \LL, \ParkS, \Katanaeveff, \Tseytlintoadd}.
However, explicit solutions and their properties have not been analysed
in detail for the specific model \tdaction\ with two arbitrary
parameters $c$ and $\Lambda$. This will be our main goal in this section.
Namely, we construct a large family of new static black hole solutions
in this model, including extremal solutions for $\Lambda < 0$ and
Schwarzschild-like black holes for $\Lambda > 0$, and discuss
their dynamical aspects, such as formation and evaporation.

\subsec{Extremal Black Hole Solution}

Variation of the action \tdaction\ with respect to the metric
and the dilaton lead to the equations of motion:
\eqn\eomg{\nabla_{\mu} \nabla_{\nu} \varphi
+ g_{\mu \nu} \Big( (\nabla \varphi)^2  - \nabla^2 \varphi
- {c \over 4} - {\Lambda \over 4} e^{2 \varphi} \Big) =0}
and:
\eqn\eomphi{R + c + 4 \nabla^2 \varphi - 4 (\nabla \varphi)^2 =0}
It is convenient to rewrite these equations of motion
in the light-cone coordinates $x^{\pp} = x^0 + x^1$,
$x^{\mm} = x^0 - x^1$, and use the conformal gauge:
$$
g_{\pp \mm} = - {1 \over 2} e^{2 \rho}, \quad
g_{\pp \pp} =0, \quad g_{\mm \mm} = 0
$$
In a compact form, this anzatse for the metric can be written as
$g_{\mu \nu} = e^{2 \rho} \eta_{\mu \nu}$,
where $\eta_{\mu \nu}$ has signature $(-,+)$.
In this gauge, the scalar curvature is
\eqn\curvconf{R = 8 e^{- 2 \rho} \partial_{\pp} \partial_{\mm} \rho ,}
the equation of motion for the metric \eomg\
becomes, {\it cf.} \refs{\CGHS,\MSW}:
\eqn\eoma{ 2 \partial_{\pp} \partial_{\mm} \varphi
- 4 \partial_{\pp} \varphi \partial_{\mm} \varphi
- {c \over 4} e^{2 \rho} - {\Lambda \over 4} e^{2(\varphi + \rho)} =0,}
and the dilaton equation of motion \eomphi\ reads:
\eqn\eomb{- 2 \partial_{\pp} \partial_{\mm} \varphi
+ 2 \partial_{\pp} \varphi \partial_{\mm} \varphi
+ \partial_{\pp} \partial_{\mm} \rho + {c \over 8} e^{2 \rho} =0.}
Since we have gauge fixed $g_{\pp \pp}$ and $g_{\mm \mm}$
to zero we must also impose their equations of motion as
constraints:
\eqn\ccca{ 2 \partial_{\pp} \rho \partial_{\pp} \varphi - \partial_{\pp}^2 \varphi
=0 }
and
\eqn\cccb{2 \partial_{\mm} \rho \partial_{\mm} \varphi - \partial_{\mm}^2 \varphi
=0}

We are interested in static solutions. Therefore, we assume that
$\rho(x)$ and $\varphi(x)$ depend only on the spatial coordinate
$x \equiv x^1 = (x_{\pp} - x_{\mm})/2$.
This gives:
\eqn\eomc{\varphi'' - 2 \varphi'^2 + {c \over 2} e^{2 \rho}
+ {\Lambda \over 2} e^{2(\varphi + \rho)} =0 }
\eqn\eomd{- 2 \varphi'' + 2 \varphi'^2 + \rho'' - {c \over 2} e^{2 \rho} =0}
\eqn\cccc{\varphi'' - 2 \varphi' \rho' =0}
for the equations of motion \eoma, \eomb,
and for the constraints \ccca\ -- \cccb, respectively.

The last equation can be easily solved to give a relation
between the conformal factor and the dilaton:
\eqn\rhoviaphi{e^{2 \rho} = A \varphi'}
for some non-zero constant $A$.
By a suitable choice of the coordinate $x$ we can always set $A=1$.
Notice that the sign of $A$ is related to the orientation of our
solution in one-dimensional space.

Substituting into \eomc\ we find:
\eqn\eome{\varphi'' - 2 \varphi'^2 = - {c \over 2} \varphi'
- {\Lambda \over 2} e^{2\varphi} \varphi' }
whose first integral is:
\eqn\fstint{\varphi' e^{-2 \varphi} - {c \over 4} e^{-2 \varphi}
+ {\Lambda \over 2} \varphi + m =0}
As we explain in the following, this solution describes
a two-dimensional black hole, and the integration
constant $m$ is proportional to the mass of the black hole.
Using the relation \rhoviaphi\ we can write down the solution for
conformal factor of the metric as a function of the dilaton:
\eqn\solrho{e^{2 \rho} = \varphi' = {c \over 4}
- e^{2 \varphi} ({\Lambda \over 2} \varphi + m) }
It can be integrated further to give an expression for the dilaton
as a function of the coordinate $x$ (which is defined up to a shift):
$$
x = -\half \int_{- \infty}^{-2 \varphi} {e^z dz \over ({c \over 4} e^z
+ {\Lambda \over 4} z -m)}
$$

At the infinity, $x \to - \infty$, the dilaton field,
given by the inverse of this function, grows linear with $x$:
\eqn\asympdil{
\varphi (x) \approx {c \over 4} x - {2 \over c} e^{{c \over 2} x}
\big({c \Lambda \over 8} x + m - {\Lambda \over 4} \big) + \ldots}
while the metric $g_{\mu \nu} \approx {c \over 4} \eta_{\mu \nu}$
is asymptotically flat. Therefore, at one spatial infinity our solution
looks like a linear dilaton background, {\it cf.} \CGHS.
It turns out that the dilaton is exactly linear in a Schwarzschild-like
gauge, where the solution \solrho\ looks like:
\eqn\schsol{ds^2 = - l(\varphi) dt^2 +  {1 \over l(\varphi)} d \varphi^2}
and the dilaton $\varphi$, which plays a role of a spatial coordinate here,
is related to the {\it tortoise} coordinate $x$ as $d \varphi = l(\varphi) dx$.
If we compare this relation with \rhoviaphi, we find that $l(\varphi)$ is
nothing but the conformal factor expressed in terms of the dilaton \solrho:
\eqn\lsol{ l(\varphi) = {c \over 4} - e^{2 \varphi} ({\Lambda \over 2} \varphi +
m) }
The Schwarzschild gauge is very convenient, so for the rest of this
section we shall work in this gauge with the solution \schsol\ -- \lsol.

Now we discuss the physical properties of this solution.
First, we compute the scalar curvature \curvconf:
$$
R = - 2 e^{- 2 \rho} \rho''
$$
which in the Schwarzschild gauge has a very simple form:
$$
R = - {d^2 l(\varphi) \over d \varphi^2} =
-2 e^{2 \varphi} (\Lambda \varphi + 2m + \Lambda)
$$
We find that $R$ goes to zero at minus infinity, $\varphi \to - \infty$,
in agreement with our earlier observation, and diverges at the plus
infinity, $\varphi \to + \infty$, where string perturbation theory breaks down.

The points in space where $l (\varphi) =0$ describe the horizon manifold:
\eqn\horzn{ {c \over 4} - e^{2 \varphi} ({\Lambda \over 2} \varphi + m) =0 }
The position and the number of such points depends on the values
of the central charge $c$ and the cosmological constant $\Lambda$.
For our solution to make sense at $\varphi \to - \infty$, where
string perturbation theory is valid, we have to assume that $c$
is positive, {\it cf.} \solrho. Then, there are two possibilities
corresponding to $\Lambda >0$ and $\Lambda < 0$, shown on Fig.1.
\ifig\pic{A plot of the conformal factor for black hole solutions
$a)$ with $\Lambda > 0$, and $b)$ with $\Lambda <0$.}
{\epsfxsize4.0in\epsfbox{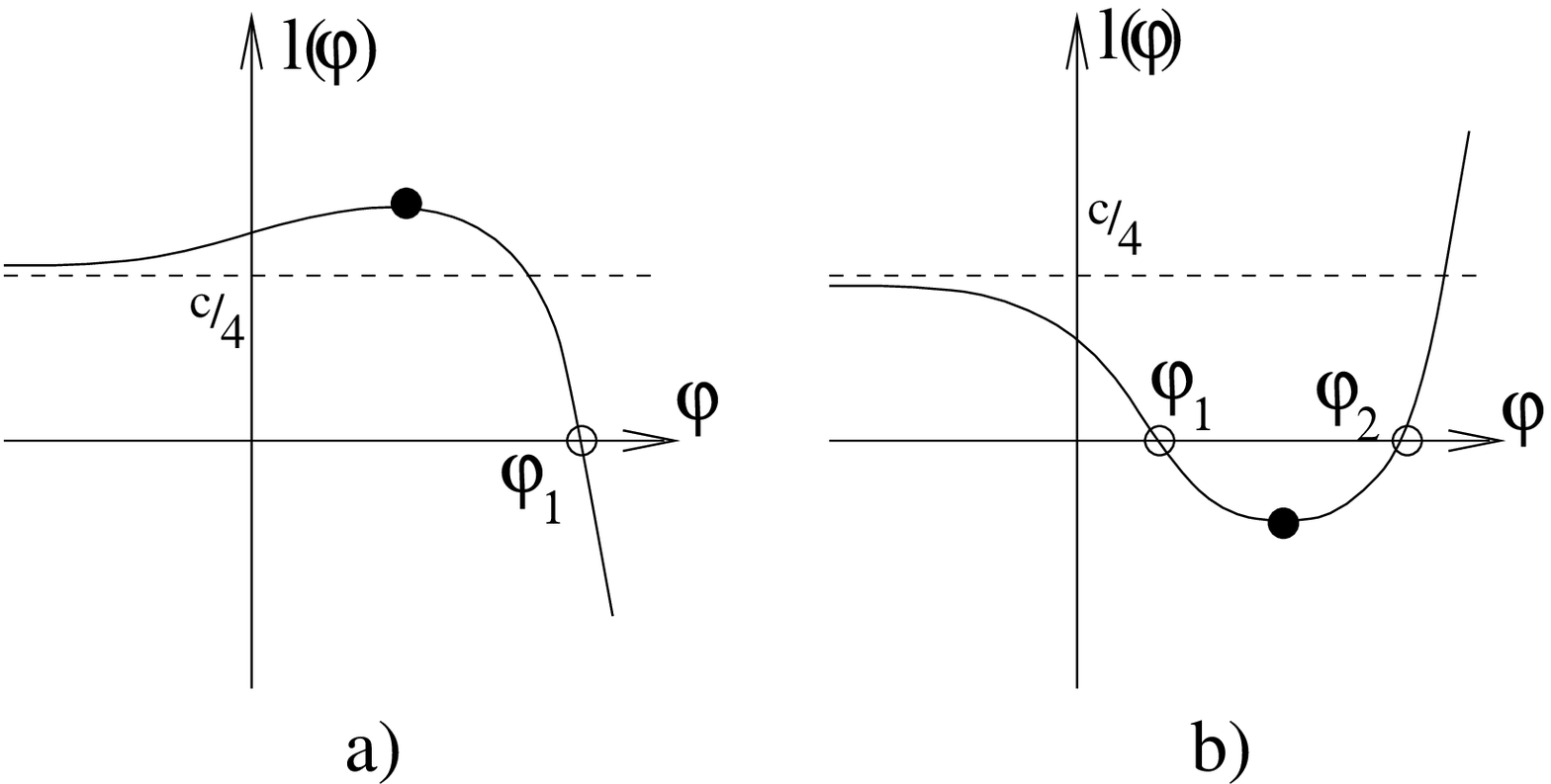}}

In the first case ($\Lambda >0$) there is always one horizon,
and the solution \lsol\ looks like a Schwarzschild black hole,
see the corresponding Penrose diagram on Fig.2a.
On the other hand, a general non-singular solution with $\Lambda < 0$
has two horizons, at $\varphi_1$ and $\varphi_2$,
just like the usual Reissner-Nordstrom black hole in four dimensions.
Similar to the Reissner-Nordstrom solution, we have
two conformal blocks for each of the intervals $(- \infty, \varphi_1)$,
$(\varphi_1, \varphi_2)$, and $(\varphi_2, + \infty)$, which altogether
represent a fundamental region for the two-dimensional analog
of the Reissner-Nordstrom black hole, see the corresponding
Penrose diagram on Fig.2b.
\ifig\pic{Penrose diagram $a)$ for a Schwarzschild-like two-dimensional
black hole ($\Lambda >0$), and $b)$ for a two-dimensional analog of
the Reissner-Nordstrom black hole ($\Lambda <0$).
Values of the dilaton are indicated at the horizons and
at the asymptotic regions.}
{\epsfxsize4.0in\epsfbox{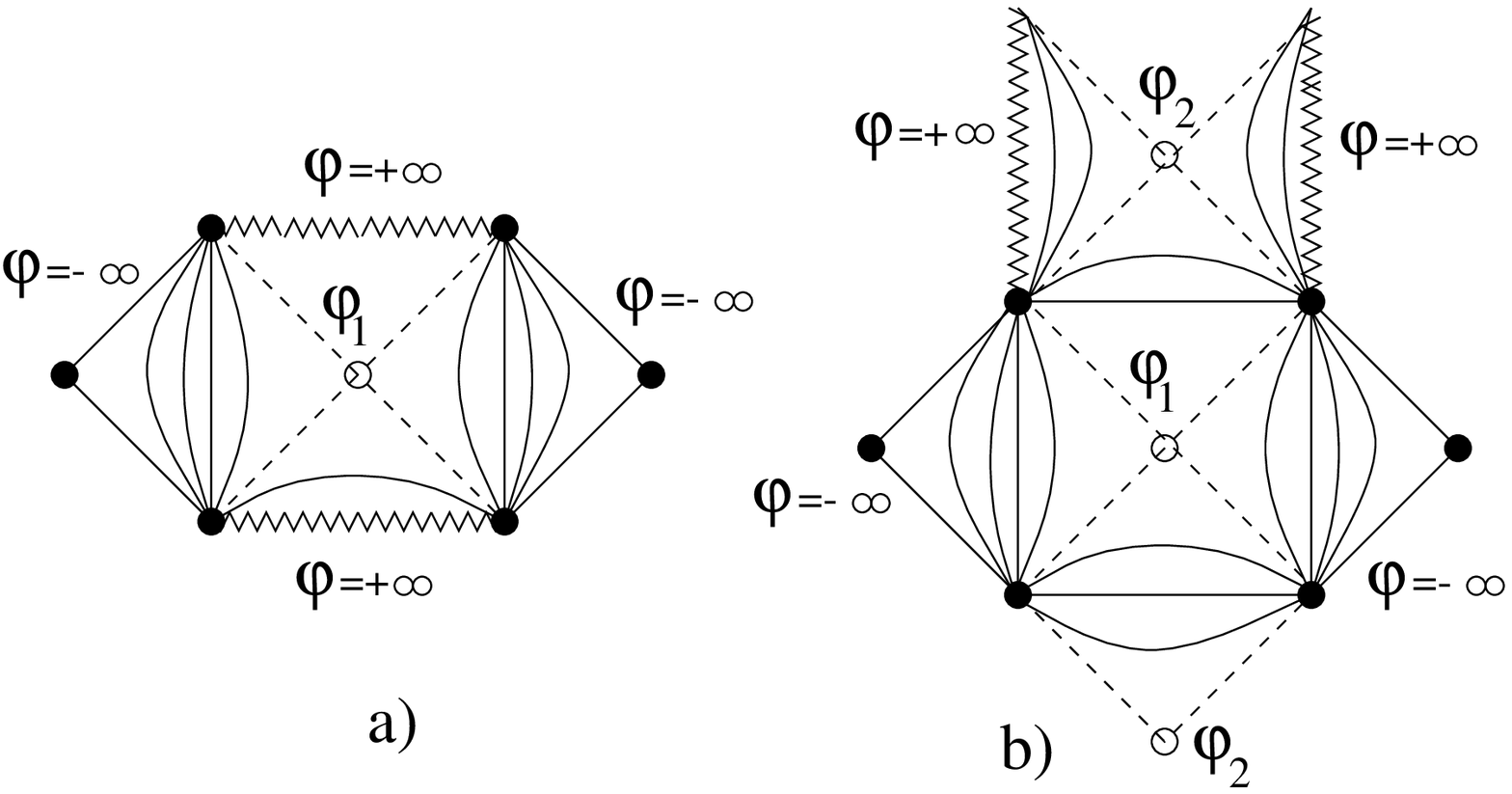}}

As we will see in a moment, extremal solutions can appear only
if $\Lambda < 0$ in agreement with our expectations from string
theory, which also suggests the existence of extremal solutions
only for negative values of the cosmological constant.
Before we demonstrate this, let us briefly comment on
the geodesic completeness across a horizon, {\it cf.} \refs{\KKL, \Katanaev}.
For a static solution the completeness of space-like geodesics
follows from the equation:
$$
\lim_{\varphi \to \varphi_h} t \to \int^{\varphi_h} 
{d \varphi \over l(\varphi)} .
$$
For our solutions \lsol, this integral always diverges at a horizon
corresponding to finite value of $\varphi$, where $l(\varphi)$ has
a simple zero (or zero of order two in the extremal case).
\ifig\pic{Conformal factor $(a)$ and the Penrose diagram $(b)$
for the extremal black hole solution. Only one fundamental region
is shown on the Penrose diagram.}
{\epsfxsize4.0in\epsfbox{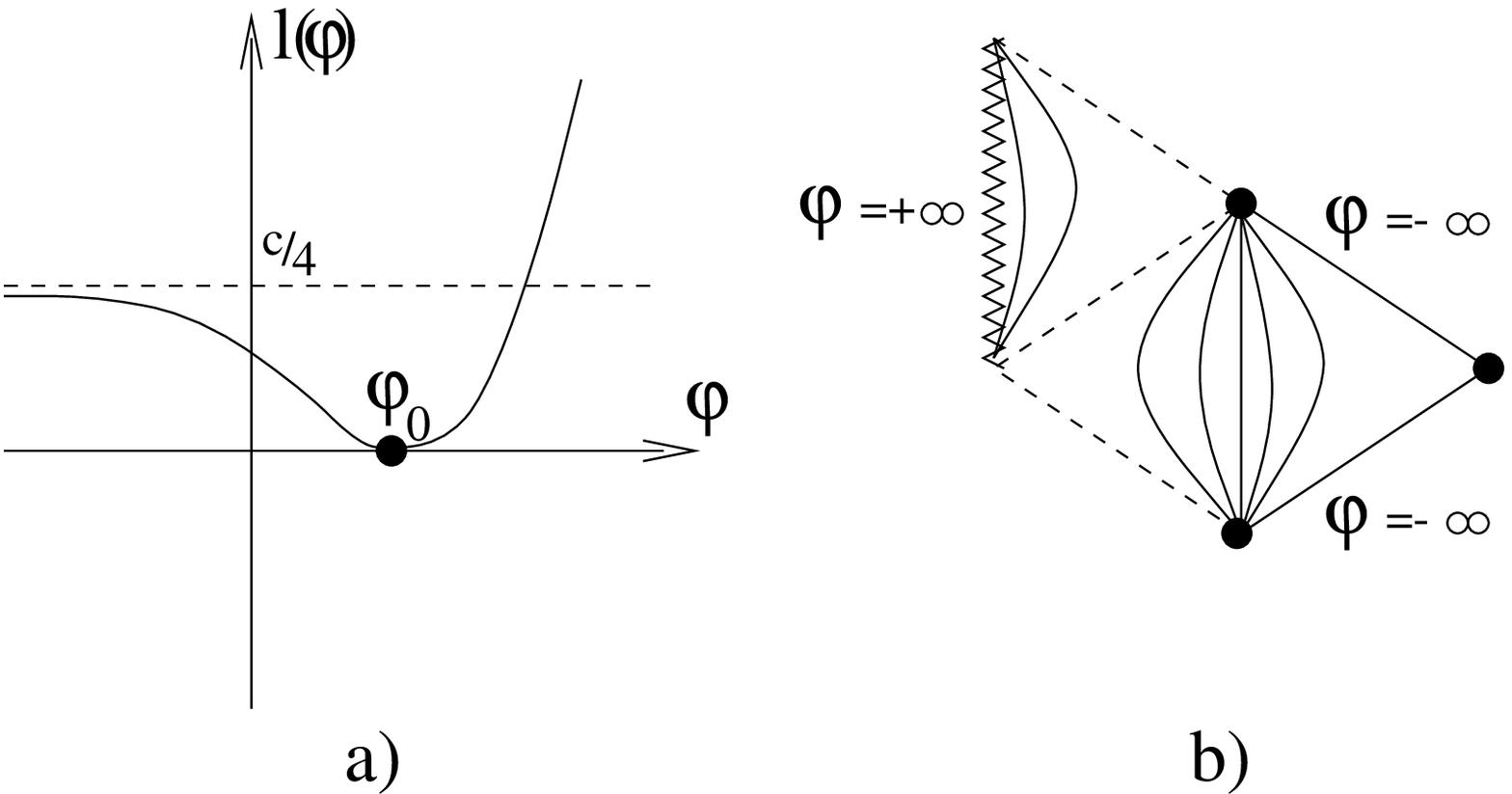}}

An extremal black hole appears when two horizons coincide,
{\it i.e.} when the function $l(\varphi)$ has a double zero
at some point $\varphi = \varphi_0$. As we mentioned above,
this may occur only for negative cosmological constant, $\Lambda <0$.
In fact, to determine the value of $\varphi_0$, we have to solve
equations: $l(\varphi)=0$ and $l' (\varphi)=0$.
Therefore, along with \horzn\ we have an equation:
\eqn\minimum{ e^{2 \varphi} (\Lambda \varphi + 2 m + {\Lambda \over 2}) =0}
which has the solution:
\eqn\mindil{\varphi_0 = - {2 m \over \Lambda} - {1 \over 2} . }
Substituting this into \horzn, we get the extremality condition
(BPS condition) for our black hole solution:
\eqn\bpsbound{m_0 =
- {\Lambda \over 4} - {\Lambda \over 4} \ln (- {c \over \Lambda}) .}
Now it is clear that for positive $c$ this expression makes sense
only if the cosmological constant $\Lambda$ is negative.
The conformal factor and Penrose diagram (that now consists of
two symmetric fundamental regions) for the extremal black hole
are shown in Fig.3.
The solution \lsol\ has a naked singularity unless the mass parameter
obeys the BPS inequality $m \ge m_0$.

We can calculate the geometric Hawking temperature as
the normal derivative of the norm of the Killing vector $\p / \p t$
at the horizon. In the Schwarzschild gauge it has a very simple form:
\eqn\thawking{ T_H = {1 \over 2} \vert {d l(\varphi) \over d \varphi} \vert_h}
{}From the explicit expression \lsol\ for the function $l(\varphi)$
it is easy to see that $T_H$ is equal to zero precisely when black becomes
extremal, {\it i.e.} when both eqns. \horzn\ and \minimum\ hold.
Therefore, in quantum theory coupled to matter, non-extremal black holes
\schsol\ -- \lsol\ are expected to emit Hawking radiation
until they reach the endpoint of evaporation -- the extremal
black hole with $m = m_0$, see \refs{\Banks} and below.
Furthermore, non-extremal solutions with $m > m_0$ are expected
to be unstable against instanton-mediated fragmentation
into disconnected universes \MMS.

On the other hand, the extremal solutions are expected to be
stable and $1/2$ BPS in the appropriate supersymmetric theory.
Although all of the black hole solutions \schsol\ constructed
here represent supersymmetric ground states in $\CN=1$ dilaton
supergravity \ParkS\ with the dilaton superpotential:
$$
\CW (\Phi) = e^{- \Phi} \sqrt{c e^{- 2\Phi} - 2 \Lambda \Phi - 4 m}
$$
only extremal solutions are expected to correspond
to BPS saturated states in the extended $\CN=2$ supergravity.
Note that in the limit $\Lambda \to 0$, the mass of the extremal solution
also goes to zero, according to \bpsbound, and the solution
degenerates into a linear dilaton background.

Note also
that in the near horizon limit, the extremal black hole solution \schsol\
looks like two-dimensional anti-de Sitter space.
For the extremal solution, $l(\varphi_0)=0$ and $l'(\varphi_0)=0$,
so that to the leading order in $(\varphi - \varphi_0)$ we have:
$$
l(\varphi) \approx {c \over 2}  (\varphi - \varphi_0)^2 .
$$
In terms of the new space-like variable
$u^{-1} = \sqrt{{c \over 2}} (\varphi - \varphi_0)$
the metric takes a more standard form:
\eqn\adstwometric{ds^2 = {1 \over u^2} (- dt^2 + R_{AdS}^2 du^2)}
where $R_{AdS} = \sqrt{{2 \over c}}$ is the characteristic scale parameter.

\subsec{Mass of the Black Hole}

Note that $m$ has dimension of mass squared,
so $m/\sqrt{c}$ is expected to be related to
the physical mass parameter up to some numerical factor.
The appropriate formula for the definition of the ADM mass in
generalized two-dimensional dilaton gravity was proposed by
Mann \Mann, and also in a recent work \LVA. Below we show that
both methods lead to the same answer for the ADM mass:
\eqn\admmass{M_{ADM} = {2 \over \sqrt{c}} m}

According to \LVA, the ADM mass of a solution like \schsol\
in a generalized dilaton gravity theory can be obtained
evaluating the following expression at the spatial
infinity\foot{The sign and normalization in this formulae
are chosen in such a way to agree with our earlier
conventions, {\it cf.} \rhoviaphi.}:
\eqn\lvaadm{M_{ADM} -M_0 = {4 \over \sqrt{c}} e^{-2 \varphi}
\Big[ \sqrt{l(\varphi) l_0 (\varphi)} - l(\varphi)
\Big]_{\varphi \to -\infty}}
where the index $0$ refers to the reference solution,
that we choose to be the extremal solution
with $m=m_0$, {\it cf.} \bpsbound.
Substituting \lsol\ into \lvaadm, we find:
\eqn\tempadm{ M_{ADM} - M_0 = {e^{-2 \varphi} \over \sqrt{c}}
\Big[\sqrt{(c - 2 e^{2 \varphi} (\Lambda\varphi + 2m_0))
(c - 2 e^{2 \varphi} (\Lambda \varphi + 2m) ) }}
$$
- (c - 2 e^{2 \varphi} (\Lambda \varphi + 2m) )
\Big]_{\varphi \to -\infty} $$
Evaluating this expression in the limit $\varphi \to - \infty$,
we get \admmass.

To provide further evidence that, up to a universal numerical factor,
the mass of the black hole is given by $m$, we use the general
formulae proposed by Mann \Mann,
that in our case reads, {\it cf.} \Nappitoadd:
\eqn\mannmass{\eqalign{ E_{ADM}
&= {1 \over 2\sqrt{c}} \Big[
e^{-2 \varphi} \big(- 4 (\nabla \varphi)^2 + c \big) + \Lambda
\Big]_{\varphi \to  \infty} \cr
&= {1 \over 2\sqrt{c}}
\Big[ e^{-2 \varphi} \big( -4 l(\varphi) + c \big) + \Lambda
\Big]_{\varphi \to  \infty} \cr
&= {1 \over 2\sqrt{c}} \Big[ 2 \Lambda \varphi + 4m + \Lambda
\Big]_{\varphi \to  \infty} }}
Once appropriately regularized, this expression clearly leads to \admmass:
$$
E_{ADM}(m) - E_{ADM}(m=m_0) = {2 \over \sqrt{c}} (m-m_0)
$$

Similar to the Reissner-Nordstrom solution, our solution \lsol\ develops
a naked singularity for $m<m_0$. Therefore, a condition that a given
black hole does not have such singularities can be expressed in
a form of the BPS inequality:
\eqn\bpsmass{M \ge M_0}
where $M_0$ is given by \bpsbound:
\eqn\emass{ M_0 =
- {\Lambda \over 2 \sqrt{c}} \Big(1 + \ln(- {c \over \Lambda}) \Big)}
Note, that for $\Lambda<0$ this expression is always positive.


\subsec{The Limit $\Lambda \to 0$}

In the limit $\Lambda \to 0$, the action \tdaction\ looks like
the usual string-inspired dilaton gravity action:
\eqn\ordinaryaction{L = \int d^{2} x \sqrt{- g} e^{-2 \varphi}
\big( R + 4 (\nabla \varphi)^2 + c \big)}
which, in the Euclidean signature, has a semi-infinite
cigar solution\foot{Our normalization for the dilaton
field in the action \tdaction\ is different from the normalization
used in \refs{\EFR,\MSW,\WittenWZW}. The notations differ by a factor
of 2; $\Phi = 2 \varphi$.} \refs{\EFR,\MSW,\WittenWZW}:
\eqn\cigards{ds^2 = {k \over 2} (dr^2 + \tanh^2 (r) d \psi^2)}
\eqn\cigardil{\varphi = - \log \cosh (r) + \varphi_0}
where $k$ is related to the central charge\foot{Note, in our
notations, $c$ is the `excess central charge,' rather than
the central charge of the $SL(2, R) / U(1)$ gauged WZW model,
denoted in \WittenWZW\ by the same letter.} $c$ as follows:
\eqn\ckrel{
c = {8 \over k-2} = {8 \over k'} }

Hence, it is natural to expect that in this limit
our solution \schsol\ looks like a cigar solution,
more precisely, like its analog in the Minkowski signature
obtained by a Wick rotation $\psi \to i t$ \WittenWZW:
\eqn\Lcigards{ds^2 = {k \over 2} (dr^2 - \tanh^2 (r) d t^2)}
Note, that the $U(1)$ isometry of the cigar solution, generated by
shifts of the angle variable $\psi$ becomes the invariance
with respect to time translations in the Minkowski signature.

Expressed in terms of the new space-like coordinate $\rho = \log \cosh (r)$
and in terms of the rescaled time-like variable $t \to (2/k)t$,
the metric \Lcigards\ indeed looks like the non-extremal
solution \schsol\ with zero cosmological constant, $\Lambda = 0$:
\eqn\LLcigards{ds^2 = - l (\rho) d t^2 + {1 \over l(\rho) } d \rho^2 }
where:
\eqn\lrho{l (\rho) = {2 \over k} (1 - e^{-2 \rho}) }
{}From \cigardil, we also find the dilaton $\varphi = - \rho + \varphi_0$,
which, in agreement with our solution \schsol, turns out to be linear.
In terms of $\varphi$, the metric \LLcigards\ is identical to our
metric \schsol\ with $\Lambda=0$ and $l(\varphi)$ given by:
$$
l(\varphi) = {2 \over k} - {2 e^{-2 \varphi_0} \over k} e^{2 \varphi}
$$
If we compare this expression with \lsol, we can identify
the parameters as follows:
\eqn\identone{ {c \over 4} = {2 \over k} }
and:
\eqn\identtwo{ m = {2 \over k} e^{-2 \varphi_0}}

The relation \identone\ between the central charge $c$
and the level $k$, in the semiclassical approximation,
is clearly compatible with the exact formula \ckrel.
Furthermore, if we start with zero cosmological constant,
$\Lambda =0$, the string one-loop amplitude (i.e. the torus partition
function) is expected to generate a non-zero $\Lambda$-term \foot
{We would like to acknowledge Steven Shenker for this comment.}.
A back reaction of this perturbation to gravity should
lead to modification of the cigar geometry \Solodukhin\
and, in particular, to bifurcation of the horizon, 
as discussed in this section.

The second relation \identtwo\ that we found comparing the two
solutions \lsol\ and \lrho\ explains the interpretation
of the parameter $m$ as the mass of the black hole.
Indeed, according to \WittenWZW, the mass of
the classical cigar solution \Lcigards\ is given by:
\eqn\massrel{
M = \sqrt{2 \over k'} e^{- 2 \varphi_0} = {2 \over \sqrt{c}} m,}
which is precisely the ADM mass formula \admmass.


\subsec{Dynamical Formation of the Near-Extremal Black Hole}

By analogy with the CGHS case \CGHS, we expect that any
perturbation of the extremal solution with critical mass $M_0$
will result in a formation of the near-extremal solution with
$M > M_0$, where $M_0$ is given by \bpsbound:
\eqn\emass{ M_0 =
- {\Lambda \over 2 \sqrt{c}} \Big(1 + \ln(- {c \over \Lambda}) \Big)}

In this subsection we explicitly describe the formation process from
interaction of the dilaton gravity with a conformal matter field $f$:
\eqn\matteraction{L = \int d^{2} x \sqrt{- g} \Big[ e^{-2 \varphi}
\big( R + 4 (\nabla \varphi)^2 + c \big) + \Lambda
- {1 \over 2} (\nabla f)^2 \Big ]}

Since the solution \lsol\ was constructed in such a way that the weak
coupling region is at $x=-\infty$, we want to set up the initial
conditions accordingly. A convenient choice of the field $f$ that
meets this requirement is where $f(x^{\mm})$ propagates from the far
past along the $x^{\pp}$ light-cone direction. To be specific,
we choose $f(x^{\mm})$  in the form of the shock wave:
\eqn\shock{{1 \over 2} (\partial_{\mm} f)^2 = m_f \delta (x^{\mm} - x_0^{\mm})}
In a similar way, we could consider an arbitrary field $f(x^{\mm})$
propagating along $x^{\pp}$ and still be able to solve the model.
It would be interesting to study integrability of the action \matteraction\
in more detail, following the ideas of \refs{\Filippov, \Zaslavskii}.

Since the matter field $f(x^{\mm})$ is assumed to be independent
on $x^{\pp}$, the constraint \ccca\ has the same form as in the theory
with $f=0$. As before, we can easily solve this constraint\foot{The numerical
factor 2 is chosen to agree with our earlier conventions.}:
\eqn\cccasol{e^{2 \rho} = 2 \partial_{\pp} \varphi}

Then, the remaining constraint and equations of motion that follow from
the modified action \matteraction\ can be written in the following
form, {\it cf.} \LVA:
\eqn\meoma{
{1 \over 2} (\partial_{\mm} \varphi) \partial_{\mm}
\Big[ \ln (\partial_{\mm} \varphi) -  \ln (\partial_{\pp} \varphi) \Big]
= {(f')^2 \over 8} e^{2 \varphi} }
\eqn\meomb{
2 \partial_{\pp} \partial_{\mm} \varphi
- 4 \partial_{\pp} \varphi \partial_{\mm} \varphi
- {c \over 2}(\partial_{\pp} \varphi)
- {\Lambda \over 2} (\partial_{\pp} \varphi) e^{2 \varphi} =0}
\eqn\meomc{
\partial_{\pp} \partial_{\mm} \Big( \varphi -
{1 \over 2} \ln (\partial_{\pp} \varphi) \Big)
= - {\Lambda \over 4}
(\partial_{\pp} \varphi) e^{2 \varphi} }

Since the last equation has the form $\partial_{\pp} (....)$,
it can be easily integrated to give:
\eqn\meomctemp{
\partial_{\mm} \Big( \varphi -
{1 \over 2} \ln (\partial_{\pp} \varphi) \Big)
= - {\Lambda \over 8} e^{2 \varphi} + \eta (x^{\mm}) }
where $\eta (x^{\mm})$ is some function of $x^{\mm}$,
which will be determined later. Integrating one more time, we get:
\eqn\meomcc{
\partial_{\mm} e^{- 2 \varphi} + {\Lambda \over 2} \varphi
+ 2 e^{- 2 \varphi} \eta (x^{\mm}) + \lambda (x^{\mm}) =0 }

We want to compare this result to the equation,
which follows from \meomb, {\it cf.} \fstint:
\eqn\meombb{
\partial_{\mm} e^{-2 \varphi} - {c \over 4} e^{-2 \varphi}
+ {\Lambda \over 2} \varphi + m (x^{\mm}) =0 }
One can immediately identify the integration functions as follows:
$$
\eta(x^{\mm}) = - {c \over 8}, \quad \lambda (x^{\mm}) = m (x^{\mm})
$$
and rewrite \meomcc\-\meombb\ as:
\eqn\meomdd{
(\partial_{\mm} \varphi) = - {c \over 8}
+ e^{2 \varphi} \Big( {\Lambda \over 4} + {m (x^{\mm}) \over 2} \Big) }
The left-hand side of this expression looks very similar to the conformal
factor $l (\varphi)$ for the black hole solution. The only difference
appears to be in the extra factor $1/2$ and in the dependence of $m$ on
the coordinate $x^{\mm}$. In fact, the extra numerical factor is needed
for \cccasol\ to agree with \lsol, and the mass of the solution we are
constructing indeed is not expected to be constant because of the shock wave.
\ifig\pic{Conformal diagram for the dynamical formation of the near-extremal
black hole from the interaction of the shock wave with the extremal solution.
The shaded regions, representing the extremal solution (before
the interaction) and non-extremal solutions (after the interaction),
are glued along the shock wave trajectory.}
{\epsfxsize4.0in\epsfbox{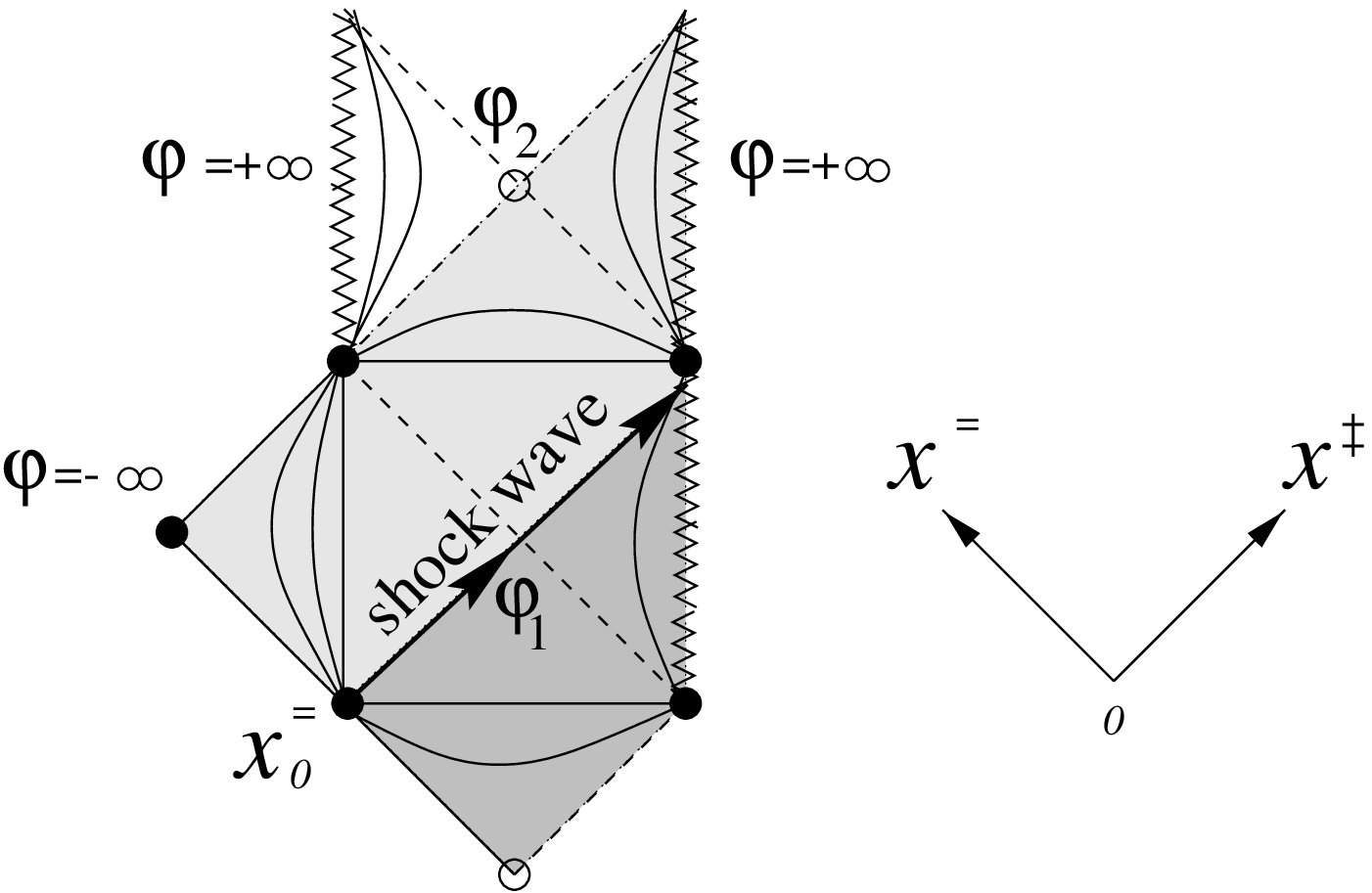}}

The evolution of the mass can be found explicitly from the remaining equation \meoma.
Using \meomctemp, we can rewrite the left-hand side of \meoma\ as:
$$
\eqalign{
{1 \over 2} (\partial_{\mm} \varphi) \partial_{\mm}
\Big[ \ln (\partial_{\mm} \varphi) -  \ln (\partial_{\pp} \varphi) \Big]
& = {1 \over 2} \partial^2_{\mm} \varphi - (\partial_{\mm} \varphi)^2 -
(\partial_{\mm} \varphi) ({\Lambda \over 8} e^{2 \varphi} - \eta) = \cr
& = {1 \over 2} \partial^2_{\mm} \varphi - (\partial_{\mm} \varphi)^2 -
(\partial_{\mm} \varphi) ({\Lambda \over 8} e^{2 \varphi} + {c \over 8}) = \cr
& = {1 \over 4} e^{2 \varphi} (\partial_{\mm} m) }
$$
Therefore, \shock\ and \meoma\ imply that $m(x^{\mm})$ is given by
the step function:
\eqn\mevol{m(x^{\mm}) = m_0 + m_f \theta(x^{\mm} - x_0^{\mm})}

Before and after $x_0^{\mm}$, the solution has the same functional
form \lsol:
$$
e^{2 \rho} = {c \over 4} - e^{2 \varphi} ({\Lambda \over 2} \varphi + m)
$$
The difference is that for $x^{\mm} < x_0^{\mm}$ the mass of the black
hole is equal to its initial value $M_0 = {2 \over \sqrt{c}} m_0$,
and after the interaction with the shock wave, $x^{\mm} > x_0^{\mm}$,
it is greater precisely by the energy of the shock wave:
$$
M = {2 \over \sqrt{c}} (m_0 + m_f)
$$

In order to glue the solutions smoothly across $x^{\mm} = x_0^{\mm}$,
one has to compute $\varphi_{<} (x^{\mm})$ before the interaction:
$$
\int^{\varphi_{<}} {d e^{- 2 \varphi} \over
({c \over 4} e^{- 2 \varphi} - {\Lambda \over 2} \varphi - m_0)  }
= x^{\mm} - x^{\pp}
$$
and evaluate this integral at $x^{\mm} = x_0^{\mm}$. This gives
the value of the dilaton field $\tilde \varphi (x^{\mm} = x_0^{\mm})$.
Substituting the result into another integral gives us a function
$h(x^{\pp})$:
$$
h(x^{\pp}) = - x_0^{\mm} + \int^{\tilde \varphi (x^{\pp})}
{d e^{- 2 \varphi} \over
({c \over 4} e^{- 2 \varphi} - {\Lambda \over 2} \varphi - m_0 - m_f)  }
$$

Now, we can write the solution $\varphi_{>} (x^{\mm})$ after
the formation of the near-extremal black hole:
$$
\int^{\varphi_{>}} {d e^{- 2 \varphi} \over
({c \over 4} e^{- 2 \varphi} - {\Lambda \over 2} \varphi - m_0 - m_f)  }
= x^{\mm} + h(x^{\pp})
$$

The classical collapse of matter coupled to dilaton gravity
is illustrated on Fig.4.


\subsec{Semiclassical Approximation}

Now we go to the next level of complication and
consider quantum effects in the limit where the Planck constant
$\hbar$ is small. We consider a generalization of the model
\matteraction\ that involves $N$ matter fields $f_i$ and
introduce a parameter $\kappa = N \hbar /12$.

A consistent approximation that takes into account
the Hawking radiation and its back reaction
on the metric was suggested by CGHS \CGHS.
It involves the limit of large $N$ with fixed $\kappa$.
In this limit
the quantum fluctuations of the dilaton and the metric are suppressed,
and one can study the quantum fluctuations of the conformal matter
semi-classically by adding to the action the trace anomaly term,
which in the conformal gauge reads:
$$
\kappa (\partial_{\pp} \rho) (\partial_{\mm} \rho)
$$
This is nothing but the one-loop Polyakov action produced by
integrating out conformal matter fields at one loop \Polyakov.
The full semi-classical action in the conformal gauge is then:
\eqn\semiaction{\eqalign{
L_{eff} =  \int d^{2} x \Big[ e^{-2 \varphi}
\big( &- 2 \partial_{\pp}\partial_{\mm} \rho
+ 4 (\partial_{\pp} \varphi)(\partial_{\mm} \varphi)
+ {c \over 4} e^{2 \rho} \big) + {\Lambda \over 4} e^{2 \rho} - \cr
& - {1 \over 2} \sum_{i=1}^N
(\partial_{\pp} f_i)(\partial_{\mm} f_i)
+ \kappa (\partial_{\pp} \rho) (\partial_{\mm} \rho) \Big ] }}

Since the matter fields are conformal and the one-loop anomaly term
does not couple to the dilaton, the equations of motion for
the dilaton and matter fields are the same as in the classical theory,
{\it cf.} \eomb:
\eqn\eomsemidil{e^{-2 \varphi} \Big(
- 2 \partial_{\pp} \partial_{\mm} \varphi
+ 2 (\partial_{\pp} \varphi) (\partial_{\mm} \varphi)
+ \partial_{\pp} \partial_{\mm} \rho + {c \over 8} e^{2 \rho} \Big) =0}
and
\eqn\eommatter{ \partial_{\pp} \partial_{\mm} f_i =0 }
The general solution to the matter field equations looks like:
\eqn\mattersoln{f_i = f_{i}^{\pp} (x^{\pp}) + f_{i}^{\mm} (x^{\mm})}

The other equations of motion and the constraint equations take the form:
\eqn\eomsemipm{
T_{\pp \mm} = e^{-2 \varphi} \Big(
2 \partial_{\pp} \partial_{\mm} \varphi
- 4 (\partial_{\pp} \varphi) (\partial_{\mm} \varphi)
- {c \over 4} e^{2 \rho} \Big)
- {\Lambda \over 4} e^{2 \rho}
- \kappa (\partial_{\pp} \partial_{\mm} \rho) =0 }
\eqn\eomsemipp{\eqalign{
T_{\pp \pp} = e^{-2 \varphi} \Big(
4 (\partial_{\pp} \varphi) (\partial_{\pp} \rho)
- 2 \partial^2_{\pp} \varphi \Big)
&+ {1 \over 2} \sum_{i=1}^N (\partial_{\pp} f_i)(\partial_{\pp} f_i)- \cr
&- \kappa \Big(
(\partial_{\pp} \rho) (\partial_{\pp} \rho)
- \partial^2_{\pp} \rho + t_{\pp} (x^{\pp}) \Big) =0 }}
\eqn\eomsemimm{\eqalign{
T_{\mm \mm} = e^{-2 \varphi} \Big(
4 (\partial_{\mm} \varphi) (\partial_{\mm} \rho)
- 2 \partial^2_{\mm} \varphi \Big)
&+ {1 \over 2} \sum_{i=1}^N (\partial_{\mm} f_i)(\partial_{\mm} f_i)- \cr
&- \kappa \Big(
(\partial_{\mm} \rho) (\partial_{\mm} \rho)
- \partial^2_{\mm} \rho + t_{\mm} (x^{\mm}) \Big) =0 }}
where two arbitrary functions $t_{\pp} (x^{\pp})$ and $t_{\mm} (x^{\mm})$
must be determined by boundary conditions, in particular, by the choice
of incoming and outgoing states. The presence of these functions is needed
in order to compensate the anomalous transformation law of the one-loop term.

In principle, one could try to study the formation and evaporation
of the near-extremal black hole in this quantum system with
the boundary conditions, for example, corresponding to
the shock wave propagating along $x^{\pp}$ at $x^{\mm} = x^{\mm}_0$,
as in the previous subsection. The partial differential equations
describing this process are \eomsemidil\ and \eomsemipm:
\eqn\ddrho{
\partial_{\pp} \partial_{\mm} \rho =
{ -2 (\partial_{\pp} \varphi) (\partial_{\mm} \varphi)
- {1 \over 8} e^{2 \rho} \big( c + 2 \Lambda e^{2 \varphi} \big)
\over \kappa e^{2 \varphi} - 1} }
and
\eqn\ddphi{
\partial_{\pp} \partial_{\mm} \varphi =
2 (\partial_{\pp} \varphi) (\partial_{\mm} \varphi)
+ {1 \over 8} e^{2 \rho} \big( c + \Lambda e^{2 \varphi} \big)
+ {\kappa \over 2} e^{2 \varphi} \partial_{\pp} \partial_{\mm} \rho }
Although hard to solve analytically, this equations can be approached
using numerical methods developed for general dilaton gravity
theories \refs{\Lowe, \LL}. We will not pursue this direction here.

Let us briefly discuss important aspects of the semi-classical
dynamics that follow from the effective Lagrangian \semiaction.
First, we note that as $\varphi \to - \infty$ the one-loop term
proportional to $\kappa$ becomes negligible (as well as
the $\Lambda$-term). Therefore, in this limit the solution
to the semi-classical action \semiaction\ should approach
the classical linear dilaton vacuum.

At the other infinity, $\varphi \to + \infty$, the semi-classical
solution is asymptotic to the de Sitter space. Indeed, in this
limit the $\Lambda$-term and the one-loop term are dominant,
and from the equation for the conformal part of the metric \eomsemipm\
we get:
$$
- {\Lambda \over 4} e^{2 \rho}
= \kappa \partial_{\pp} \partial_{\mm} \rho
$$
Substituting this into \curvconf, we find the asymptotic
value of the curvature as $\varphi \to + \infty$:
\eqn\dscurv{R \approx {- 2 \Lambda \over \kappa}
= {- 24 \Lambda \over N \hbar} }
Since $\Lambda$ is assumed to be negative (for otherwise we had
no extremal solution), locally we get a de Sitter space.
It is curious to note that this de Sitter space is generated
entirely by quantum effects (some other interesting quantum
phenomena can be found in \Odintsov). Therefore, it is natural to expect
that its entropy, proportional to $N$, can be microscopically
understood as quantum entaglement with degrees of freedom
hidden behind the horizon \ENTANGLE.
It would be interesting to verify this by a direct calculation,
analogous to \HMS.

Note also, that the curvature \dscurv\ of this semiclassical
de Sitter space is inversely proportional to the number, $N$,
of matter fields. This relation is reminiscent of the $\Lambda - N$
correspondence proposed by T.~Banks \BanksN, although one should
remember that the model we are considering has infinite-dimensional
Hilbert space (in fact, each scalar field $f_i$ has infinite-dimensional
Hilbert space.)

The strong coupling region is described by the Liouville gravity
with positive cosmological constant and conformal matter:
$$
L_{eff} (\varphi \to + \infty) \approx \int d^{2} x \Big[
{\Lambda \over 4} e^{2 \rho} - {1 \over 2} \sum_{i=1}^N
(\partial_{\pp} f_i)(\partial_{\mm} f_i)
+ \kappa (\partial_{\pp} \rho) (\partial_{\mm} \rho) \Big ]
$$
Since this theory has massive excitations, one might expect that
a shock wave propagating along $x^{\pp}$ will radiate part of its
energy to $\CI^{\pp}_L = \{ x^{\mm} \to + \infty \}$, but part
can be carried into the Liouville region, so one might expect
some stable remnants of the black hole evaporation,
unlike the CGHS case \CGHS.

However, this part of space is separated from the weak coupling
region by a curvature singularity at:
\eqn\singularity{ e^{- 2 \varphi} = \kappa}
This strong coupling singularity is actually present also in
the CGHS solution and becomes visible at some stage of the black
hole evaporation \Hawking. The reason is that CGHS solution has
finite evaporation rate when the dilaton field at the horizon
approaches the critical value \singularity. Our black hole solution
would have the same fate, unless the rate of radiation goes to
zero at some point. If this happens, the singularity can be hidden behind
the event horizon.
Below we present some arguments that this is indeed the case.


\subsec{Evaporation of the Near-Extremal Black Hole}

Our original classical Lagrangian \tdaction\ is a specific example
of a non-singular Lagrangians studied by Banks and O'Loughlin \Banks.
The authors of \Banks\ argued that, in general, the Lagrangians like
\tdaction\ are expected to have extremal solutions, which are nonsingular
analogues of the extremal Reissner-Nordstrom solutions, and moreover
that these solutions are the endpoint of the Hawking evaporation
when the models are coupled to matter. Here we present some arguments
that the black hole solutions \schsol\ studied here, which saturate
the extremality condition \bpsbound\ are indeed the terminal points
of the evaporation of non-extremal solutions produced, say,
via collapse of an $f$-wave, see Fig.4.
A similar problem of a charged gravitational collapse with
the full nonlinear semiclassical back-reaction was studied
recently in \Sorkintoadd.

First, let us neglect the back reaction of the Hawking radiation.
Then, the evaporation rate is given by the value of the energy
flux across $\CI^{\pp}_L$ at the horizon.
Since the black hole solution \lsol\ obeys the classical equations
of motion, the first line in \eomsemipp\ is zero, and we get:
\eqn\fluxpp{\langle T_{\pp \pp} \rangle =
- \kappa \Big( (\partial_{\pp} \rho) (\partial_{\pp} \rho)
- \partial^2_{\pp} \rho \Big) }
The right-hand side of this expression is supposed to be evaluated
for the classical solution \lsol. In order to do this, we use
\solrho\ to express the derivatives of $\rho$ in terms of $l (\varphi)$:
$$
\partial_{\pp} \rho
= {1 \over 4 l} {d l \over d \varphi} {d \varphi \over d x}
= {1 \over 4} {d l \over d \varphi}
$$
$$
\partial^2_{\pp} \rho
= {1 \over 8} {d \over d x} \Big({d l \over d \varphi} \Big)
= {1 \over 8} l {d^2 l \over d \varphi^2}
$$
Therefore, we can write \fluxpp\ as \KKL:
\eqn\fluxppp{\langle T_{\pp \pp} \rangle =
- {\kappa \over 8} (2 (l')^2 - l'' l)}
To get the evaporation rate we have to evaluate this expression
at the horizon, where $l (\varphi) =0$:
\eqn\fluxpppp{\langle T_{\pp \pp} \rangle_h
=- {\kappa \over 4} (l')^2 = - {\kappa \over 4}
e^{4 \varphi} (\Lambda \varphi + 2m + {\Lambda \over 2} )^2 }
If we compare the right-hand side of this expression with the formula
\thawking\ for the geometric Hawking temperature, we find \KKL:
$$
\langle T_{\pp \pp} \rangle_h =- {\kappa \over 4} T_H^2
$$
It follows that the evaporation rate is zero precisely for the extremal
black hole, {\it i.e.} when both
$l (\varphi)$ and $l' (\varphi)$ vanish at the horizon.
Therefore, the extremal solution \lsol\ with the critical mass \bpsbound\
is expected to be a stable remnant of the evaporation of non-extremal
black hole. Quantum fields in this gravitational background will not
Hawking radiate.

To provide a further evidence for this picture, we consider adiabatic
evaporation of a black hole with the mass slightly above extremal:
$$
m > m_0 =
- {\Lambda \over 4} - {\Lambda \over 4} \ln (- {c \over \Lambda})
$$
For example, this black hole can be produced in the collapse of
the shock wave, as we discussed above.

In the rest of this section we use the light-cone gauge, which
turns out to be more convenient for this calculation. For example,
the black hole metric in this gauge looks like:
\eqn\lcmetric{ds^2 = 2 dv dr + l(r) dv^2, \quad \varphi=r }
and the shock wave is described by
$\epsilon^{\mu} \epsilon^{\nu} T_{\mu \nu}^f = 2 m_f \delta(v - v_0)$,
where $\epsilon = (l/2,1)$ is a null vector.
The scalar curvature in the light-cone gauge is given simply
by the second derivative of the function $l$.

Following \refs{\ST,\LL}, we introduce:
\eqn\sigdef{\Sigma = 2 \partial_v \varphi + l \partial_r \varphi}
so that a horizon is given by $\Sigma=0$.
Before the interaction with the shock wave, the space-time is described
by the extremal solution \lsol\ with the critical mass parameter $m_0$
and the apparent horizon at \mindil:
$$
\varphi_0 = {1 \over 2} \ln \big( - {c \over \Lambda} \big)
$$
After the interaction takes place, the horizon splits into "inner"
and "outer" horizons, located at $\varphi_+$ and $\varphi_-$.
Let us compute $\varphi_{\pm}$ in the approximation $m_f \ll \Lambda$.
In this approximation, the values of $\varphi_{\pm}$ are supposed
to be close to the original position of the horizon, $\varphi_0$.

Since both $l(\varphi_0)$ and $l'(\varphi_0)$ vanish for the extremal
solution, from \sigdef\ we have:
\eqn\delhor{ \Sigma (\varphi \approx \varphi_0)
= {1 \over 2}  (\varphi - \varphi_0)^2 l''(\varphi_0)
= {c \over 2}  (\varphi - \varphi_0)^2 }
On the other hand, the discontinuity of $\Sigma$ across
the shock wave is given by \LL:
\eqn\delsig{ \delta \Sigma
= {2 m_f \exp (2 \varphi) \over \sqrt{1 - \kappa \exp (2 \varphi) } }
\Big|_{\varphi = \varphi_0} \approx - 2 m_f {c \over \Lambda} }
The new positions of the horizon are given by $\Sigma + \delta \Sigma =0$.
Therefore, substituting \delsig\ into \delhor, we obtain:
\eqn\newhor{\varphi_{\pm} = \varphi_0 \pm \sqrt{ - {4m_f \over \Lambda}} }
Furthermore, from \delhor\ it follows that:
\eqn\dphisig{\partial_{\varphi} \Sigma (\varphi_{\pm})
\approx \pm (\varphi_{\pm} - \varphi_0) l''(\varphi_{\pm})
= \pm c \sqrt{ - {4m_f \over \Lambda}} }

In order to find the adiabatic evolution of the $\varphi_{\pm}$,
we also need to find $\partial_{v} \Sigma (\varphi_{\pm})$.
This can be obtained from the following linear combination of
the constraint equations \eomsemipm\ - \eomsemimm:
\eqn\lcconstr{\eqalign{
\epsilon^{\mu} \epsilon^{\nu} T_{\mu \nu}
&= {\kappa \over 4} \big({l \over 2} \partial_{\varphi}^2 l
+ \partial_{v} \partial_{\varphi} l
- {1 \over 4} (\partial_{\varphi} l)^2 \big) = \cr
&= e^{- 2 \varphi} \big(
\partial_v \Sigma + {l \over 2} \partial_{\varphi} \Sigma
- {1 \over 2} \Sigma^2 - {1 \over 2} \partial_{\varphi} l \Sigma \big) }}
Here, we took into account that the near-extremal solution
obeys classical constraint equations, and also that $t_{\varphi}$
is zero for static solutions with no net flux at infinity \LL.
If we evaluate this constraint equation for $\varphi = \varphi_{\pm}$,
the leading order terms for small $m_f$ give:
\eqn\dvsig{\partial_v \Sigma (\varphi_{\pm}) =
{\kappa \over 4} e^{2 \varphi} {\delta l \over 2} l'' }
The discontinuity of $l (\varphi)$ across the shock wave is given by \LL:
$$
\delta l = {8 m_f \over \kappa} \int d \varphi
\Big(-1 + {1 \over \sqrt{1 - \kappa \exp(2 \varphi)}} \Big)
\approx 2 m_f e^{2 \varphi}
$$
If we substitute this into \dvsig\ and evaluate the result at the new
horizon $\varphi = \varphi_{\pm}$, we get:
\eqn\dvsigma{\partial_v \Sigma (\varphi_{\pm}) = {\kappa m_f c^3 \over 2 \Lambda^2} }

Now we are ready to put things together, and compute the relative
positions of the horizons as a function of $v$.
In fact, in the adiabatic approximation we can write:
\eqn\dvphipm{\partial_v \varphi_{\pm}
= - {\partial_v \Sigma \over \partial_{\varphi} \Sigma}
= {\kappa c^2 \over 8 \Lambda} (\varphi_{\pm} - \varphi_0) }
The approximation we are using here holds if positions of the horizons
are moving slowly enough. This implies the consistency condition
$\kappa c^2/ \Lambda^2 \ll 1$. When this condition is satisfied, we can
integrate \dvphipm\ with the result:
\eqn\slowsol{\varphi_{\pm} = \varphi_0
\pm (\delta \varphi_0) e^{{\kappa c^2 \over 8 \Lambda}(v - v_0)} }
This means that the black hole mass goes back to the extremal value as:
\eqn\slowmass{M(v)
= M_0 + M_f e^{{\kappa c^2 \over 4 \Lambda}(v - v_0)} }
The Penrose diagram in the coordinates $(v,\varphi)$ for this evaporation
process is almost identical to the corresponding diagram for the evaporating
Reissner-Nordstrom black hole \ST.

\newsec{Discussion}

In this paper we proposed a hybrid formalism for the manifestly
spacetime
supersymmetric quantization of the superstring in
curved two-dimensional backgrounds with Ramond-Ramond flux,
thus extending the list of existing covariant descriptions
of superstrings in four \Nat, six \refs{\NV,\NVW},
and ten \pures\ space-time dimensions.
Such two-dimensional backgrounds appear, for example, in Type IIA
string theory on Calabi-Yau four-folds, where R-R flux is required
in general for 
the global anomaly cancellation condition \refs{\SVW, \Wflux}:
\eqn\sethrel{{\chi\over 24} =
N + {1\over 2 (2\pi)^2}\int_X F^{(4)} \wedge F^{(4)}}
Here $\chi (X)$ is the Euler number of the Calabi-Yau four-fold $X$,
$N$ is the number of fundamental strings filling two-dimensional space-time,
and $F^{(4)}$ is the value of the background Ramond-Ramond 4-form flux.
So, when $N\neq {\chi\over {24}}$,
this consistency condition requires non-zero R-R flux.
One might hope to study certain
aspects of these models using the covariant approach, as in section 2.7.

Another interesting example of curved two-dimensional backgrounds
discussed in this paper is a new black hole solution constructed
in section 4. This application is not expected to be related to
CY four-fold compactifications, since the latter leads to
an effective action \tdaction\ with zero $c$-term,
at least to the leading order in $\a '$.
We expect, however, that the black hole solution of \schsol\ -- \lsol\
can be described by superstring theory on some supergroup
manifold, in analogy with the $SL(2,R)/U(1)$ gauged WZW model of
\WittenWZW\ for the $\Lambda=0$ black hole solutions.

In principle, one can obtain this supergroup manifold (to leading
order in $\a '$) by plugging the black hole solution of 
\schsol\ -- \lsol\ into the background superfields which appear
in the sigma model action of \two\ and \ft. However, since the 
solution is rather complicated (e.g. it contains a non-constant
dilaton), it may be simpler to try to guess the appropriate supergroup.
One natural guess for the supergroup manifold is a gauged WZW
model based on the supergroup
%
\eqn\coset{G/H = OSp(1,1 \vert 2) / SO(1,1) \times SO(2).}
Indeed, this coset supermanifold has the right dimension to
describe an $N=(2,2)$ target superspace,
{\it viz.} two bosonic and four fermionic generators.
However, the WZW model based on this supergroup
can be shown to be equivalent to the bosonic $SL(2,R)$ WZW model
plus a set of free fermions \bars. This result is similar to the
hybrid version of the sigma model action for an $AdS_3\times S^3$
background with pure NS-NS flux \NVW\ where the $PSU(2\vert 2)$ WZW
model is equivalent to the bosonic $SU(2)\times SU(1,1)$ WZW
model plus a set of free fermions.

So in analogy with the $AdS_3\times S^3$ sigma model, one does not
expect the gauged WZW model based on the supergroup of \coset\ to
describe solutions with R-R flux. However, in the $AdS_3\times S^3$
case, it was possible to introduce R-R flux by including additional
$PSU(2\vert 2)$-invariant terms in the action which spoil the holomorphic
structure of the WZW currents but do not break the worldsheet
$N=(2,2)$ superconformal invariance. So it is reasonable to expect
that one can include R-R flux in the black hole solutions
by introducing similar $OSp(1,1\vert 2)$-invariant terms into
the gauged WZW model based on the supergroup of \coset. It
would be very useful to find the explicit form of such terms and
a preliminary search has already begun \toappear.
One might also hope to find a matrix model for the black
hole \schsol\ -- \lsol\ in an R-R background,
which in the limit $\Lambda \to 0$ would reproduce
the matrix model of Kazakov, Kostov, and Kutasov \KKK.

\centerline{\bf Acknowledgments}

We are grateful to C.~G.~Callan, O.~Chand\'{\i}a,
A.T.~Filippov, G.T.~Horowitz, J.~Maldacena, D.~Nedel,
R.~Plesser,  H.~Ooguri, V.~Rivelles, J.H.~Schwarz,
N.~Seiberg, S.~Shenker, A.~Strominger, C.~Vafa,
and E.~Witten for useful discussions and comments.
The work of N.B. is supported in part by
CNPq grant 300256/94-9, Pronex 66.2002/1998-9,
and FAPESP grant 99/12763-0.
This research was partially conducted during the period N.B.
served as a Clay Mathematics Institute Prize Fellow and S.G.
served as a Clay Mathematics Institute Long-Term Prize Fellow.
The work of S.G. is also supported in part
by the Caltech Discovery Fund, NSF grant No. PHY99-07949, grant RFBR
No. 01-02-17488, and the Russian President's grant No. 00-15-99296.
The work of B.C.V. is supported by FAPESP grant No 00/02230-3.
N.B. and S.G. are grateful to the California Institute of Technology,
where part of this work was done, for hospitality.


\listrefs
\end